\documentclass[]{IEEEtran}
\usepackage{cite}
\usepackage{amsmath,amssymb,amsfonts}
\usepackage{algorithmic}
\usepackage{graphicx}
\usepackage{textcomp}
\usepackage{xcolor}
\usepackage{subfig}
\usepackage{tabularx}
\usepackage{array}
\usepackage{booktabs}
\usepackage{hyperref}
\usepackage{bm}
\usepackage{xspace}
\newcommand{\bmath}[1]{\ensuremath{\bm{#1}}\xspace}

\newcommand{\x}{\bmath{x}}
\newcommand{\y}{\bmath{y}}
\newcommand{\z}{\bmath{z}}

\newcommand{\f}{\bmath{f}}

\newcommand{\rv}{\bmath{r}}
\newcommand{\w}{\bmath{w}}

\newcommand{\1}{\bmath{1}}
\newcommand{\alp}{\bmath{\alpha}}
\newcommand{\bet}{\bmath{\beta}}
\newcommand{\tht}{\bmath{\theta}}

\newcommand{\muv}{\bmath{\mu}}

\newcommand{\A}{\bmath{A}}

\newcommand{\I}{\bmath{I}}
\newcommand{\K}{\bmath{K}}

\newcommand{\Pm}{\bmath{P}}

\newcommand{\beq}{\begin{equation}}
\newcommand{\eeq}{\end{equation}}
\newcommand{\bea}{\begin{eqnarray}}
\newcommand{\eea}{\end{eqnarray}}
\newcommand{\ba}{\left(\!\!\begin{array}}
\newcommand{\ea}{\end{array}\!\!\right)}
\newcommand{\bc}{\begin{center}}
\newcommand{\ec}{\end{center}}

\def\BibTeX{{\rm B\kern-.05em{\sc i\kern-.025em b}\kern-.08em
		T\kern-.1667em\lower.7ex\hbox{E}\kern-.125emX}}

\newcommand{\txtc}[1]{\textcolor{black}{#1}}
\newcommand{\txtb}[1]{\textcolor{black}{#1}}

\begin{document}
\title{{Neural KEM: A Kernel Method with Deep Coefficient Prior for PET Image Reconstruction}}
\author{Siqi Li, Kuang Gong,  Ramsey D. Badawi, Edward J. Kim, Jinyi Qi,  and Guobao Wang
	
\thanks{This work is supported in part by an investigator-initiated research grant from United Imaging Healthcare of America and by NIH under grant no. R01DK124803. Part of this paper was presented in the 2021 IEEE Nuclear Science Symposium and Medical Imaging Conference. }
\thanks{S. Q. Li is with the Department of Radiology, University of California Davis Health, Sacramento, CA 95817, USA. (e-mail: sqlli@ucdavis.edu).}
\thanks{K. Gong is with the Gordon Center for Medical Imaging, Massachusetts General Hospital and Harvard Medical School, Boston, MA 02114, USA. (e-mail: KGONG@mgh.harvard.edu)}
\thanks{R. D. Badawi is with the Department of Radiology, University of California Davis Health, Sacramento, CA 95817, USA. (e-mail: rdbadawi@ucdavis.edu)}
\thanks{E. J. Kim is with the Comprehensive Cancer Center, University of California Davis Health, Sacramento, CA 95817, USA. (e-mail: jhkim@ucdavis.edu)}
\thanks{J. Qi is with the Department of Biomedical Engineering, University of California at Davis, Davis, CA 95616, USA. (e-mail: qi@ucdavis.edu)}
\thanks{G. B. Wang is with the Department of Radiology, University of California Davis Health, Sacramento, CA 95817, USA. (e-mail: gbwang@ucdavis.edu).}}

\maketitle

\begin{abstract}
Image reconstruction of low-count positron emission tomography (PET) data is challenging. Kernel methods address the challenge by incorporating image prior information in the forward model of iterative PET image reconstruction.  The  kernelized expectation-maximization (KEM) algorithm has been developed and demonstrated to be effective and easy to implement. A common approach for a further improvement of the kernel method would be adding an explicit regularization, which however leads to a complex optimization problem. 
In this paper, we propose an implicit regularization for the kernel method by using a deep coefficient prior, which represents the kernel coefficient image in the PET forward model using a convolutional neural-network. To solve the maximum-likelihood neural network-based reconstruction problem, we apply the principle of optimization transfer to derive a neural KEM algorithm. Each iteration of the algorithm consists of two separate steps: a KEM step for image update from the projection data and a deep-learning step in the image domain for updating the kernel coefficient image using the neural network. This optimization algorithm is guaranteed to monotonically increase the data likelihood.
The results from computer simulations and real patient data have demonstrated that the neural KEM can outperform existing KEM and deep image prior methods.
\end{abstract}

\section{Introduction}

\IEEEPARstart{T}{omographic} image reconstruction for positron emission tomography (PET) is challenging because of the ill-conditioned problem and low counting statistics \cite{Qi2006}. 
The kernel method addresses this challenge by integrating image prior information into the forward model of PET image reconstruction (e.g., \cite{Wang2015, Hutchcroft2016, Novosad2016, Wang2019, Bland2018, Gong2018, Deidda2019, Deidda2019a}). The {\em a priori} information can come from composite time frames in a dynamic PET scan \cite{Wang2015, Wang2019}, or from the co-registered anatomical images, e.g., magnetic resonance (MR) images \cite{Hutchcroft2016, Novosad2016}. The derived kernel expectation-maximization (KEM) algorithm \cite{Wang2015} has been demonstrated to be effective and is easy to implement \cite{Wang2015, Hutchcroft2016, Novosad2016, Wang2019, Zhang2020}. 

To further improve the kernel method such as for higher temporal resolution dynamic PET imaging or for low-dose PET imaging, a straightforward approach would be adding an explicit regularization form on the kernel coefficient image to stabilize the solution \cite{Wang2015}. This can be achieved using either conventional penalty functions  (e.g., \cite{Bowsher1996, Chan2009, Cheng-Liao2011, Vunckx2013}) or convolutional neural network (CNN) based penalties (e.g., \cite{Kim2018, XieN2020}). However, such regularization-based methods generally require a complex optimization algorithm, involve one or more hard-to-tune hyper-parameters, and need to run for many iterations for a \txtb{convergent solution.} 

In this paper, we propose an implicit regularization for the kernel method by using CNN to represent the kernel coefficient image in the kernelized model.  The use of CNN representation shares the same spirit of the work of Gong {\em et al.} \cite{Gong2019, Gong2019a, Gong2021} and others \cite{Xie2020, Yokota2019, Xie2021} that employs deep image prior (DIP) \cite{Ulyanov2020} for PET image reconstruction. Differently the CNN representation in this work is applied in the kernel coefficient space instead of the original PET activity image space, resulting in a modified kernel method with deep coefficient prior. 

One challenge with solving the corresponding optimization problem is that the neural network is involved in the projection domain, resulting in a large-scale, nonlinear reconstruction problem. The alternating direction method of multipliers (ADMM) is a popular optimization approach to solving this kind of problems,  e.g.,  in \cite{Gong2019, Xie2020}. However, the hyper parameters associated with an ADMM algorithm are challenging to tune in practice.  In this work, we derive an easy-to-implement iterative algorithm by using the principle of optimization transfer \cite{Lange2000, Fessler2010, Wang2012} for the neural network-based reconstruction.  We call the new algorithm {\em neural} KEM to differentiate it from the original KEM algorithm.  

There are also other ways to explore deep learning for PET reconstruction \cite{Reader2021, Gong2020, Wang2020, Ge2020, Ge2021}, such as the direct end-to-end mapping of PET image from projection \cite{Haggstrom2019} and unrolled model-based deep-learning reconstruction \cite{Lim2020, Mehranian2020}. All these approaches require pre-training using a large population-based database, which is not always available. Similar to the original kernel method \cite{Wang2015} and the DIP method \cite{Gong2019}, the proposed method does not require population-based pretraining but is solely based on the data of single subjects. 

The remaining of this paper is organized as follows. Section II introduces the background materials of the kernel method and DIP method for PET image reconstruction. Section III describes the proposed neural KEM method that combines the kernel method with deep coefficient prior. We then present 2D and 3D computer simulation studies in \txtb{Section IV and V} and a real patient data study in Section VI to demonstrate the improvement of the proposed method over existing methods. Finally discussions and conclusions are drawn in Section VII and VIII.
\section{Background}
\subsection{PET Image Reconstruction}
PET projection measurement $\y = \left\{y_{i}\right\}_{i=1}^N$ can be well modeled as independent Poisson random variables using the log-likelihood function \cite{Qi2006},
\beq
L(\y|\x) = \sum_{i=1}^{N}y_{i} \log \overline{y}_{i} - \overline{y}_{i} - \log y_{i}!,
\eeq
where the expectation of the projection data, $\overline{\y} = \left\{\overline{y}_i\right\}_{i=1}^N$, is related to the unknown activity image $\x = \left\{x_{j}\right\}_{j=1}^J$ through
\beq
\overline{\y} = \Pm \x + \rv,
\label{PET model}
\eeq
where $x_{j}$ denotes the PET image intensity value in pixel $j$. $N$ is the total number of detector pairs and $J$ is the number of image pixels. $\Pm $ is the detection probability matrix and includes normalization factors for scanner sensitivity, scan duration, deadtime correction and attenuation correction. $\rv$ is the expectation of random and scattered events\cite{Qi2006}.

The maximum likelihood estimate of the activity image $\x$ is found by maximizing the Poisson log-likelihood,
\beq
\hat{\x} =\arg\max\limits_{\x \ge 0}L(\y|\x).
\label{ML}
\eeq
A common way of seeking the solution of (\ref{ML}) is the maximum likelihood expectation maximization (ML-EM) algorithm \cite{Shepp1982}.

\subsection{Kernel EM for PET Reconstruction}

The image estimate by standard MLEM is commonly noisy due to the limited counting statistics of PET emission data. To suppress noise, the kernel method\cite{Wang2015} incorporates an image prior into the forward projection of PET reconstruction by describing the image intensity $x_j$ using kernels,
\beq
x_j = \sum_{l\in\mathcal{N}_j} \alp_l \kappa(\f_j,\f_l),
\label{kernel representation}
\eeq
where $\mathcal{N}_j$ defines the neighborhood of pixel $j$, e.g., by the k-nearest neighbors (kNN, \cite{Friedman1977}). $\kappa(\cdot,\cdot)$ is the kernel function (e.g., radial Gaussian) and $\f$ denotes the low-dimensional feature vector that is extracted at each pixel from the image prior $\z$ (e.g. the composite images in dynamic PET or anatomical image in PET/MR or PET/CT).
The equivalent matrix-vector form of (\ref{kernel representation}) is
\beq
\x = \K \alp,
\label{kernel model}
\eeq
where $\K$ is a sparse square kernel matrix with its ($j,l$)th element being $\kappa(\f_j,\f_l)$. $\alp$ denotes the corresponding kernel coefficient image.

Substituting the kernelized image model (\ref{kernel model}) into the standard PET forward projection model in (\ref{PET model}) gives the following kernelized forward projection model for PET image reconstruction,
\beq
\overline{\y} = \Pm\K\alp + \rv.
\label{eq-kernel}
\eeq

The maximium-likelihood estimate of $\alp$ can be found by the kernel EM algorithm \cite{Wang2015},
\beq
\alp^{n+1} = \frac{\alp^n}{\w}\cdot\left( \K^T\Pm^T\frac{\y}{\Pm\K\alp^n+\rv}\right),
\label{KEM}
\eeq
where 
\beq
\w=\K^T\Pm^T\1_N,
\label{eq-w}
\eeq
and $\1_{N}$ is a vector with all elements being 1. $n$ denotes the iteration number and the superscript ``$T$'' denotes matrix transpose. The vector multiplication and division are element-wise operations. Note that the KEM update becomes the standard EM update if $\K$ is an identity matrix.
Once $\alp$ is estimated, the final reconstructed PET image is given by
\beq
\hat{\x} = \K\hat{\alp}.
\eeq
\txtb{Note that the same update equation (\ref{KEM}) is also used by the sieves method \cite{Snyder1985}. The difference is that the sieves method uses a stationary Gaussian kernel \cite{Snyder1985}, while the kernel method here uses data-driven spatially variant kernels that are derived from the image prior $\z$.}

The estimated kernel coefficient image $\alp$ by the standard kernel method \cite{Wang2015} may still suffer from noise, as demonstrated later in this paper. One possible way for improvement is to add an explicit penalty function to stabilize the estimation of $\alp$ as indicated in the original kernel method \cite{Wang2015}, which however may result in a more challenging optimization problem and involves at least one more regularization parameter to tune.

\subsection{PET Reconstruction Using DIP}
The DIP method for PET reconstruction is proposed in \cite{Gong2019} based on the representation ability of CNNs. Instead of using the linear kernel representation in (\ref{eq-kernel}), the PET image $\x$ can be also described by a nonlinear representation using neural networks and the image prior data $\z$,
\beq
\x = \bet(\tht|\z),
\label{DIP}
\eeq
where $\bet$ is a neural netwok model with $\z$ the input  images and $\tht$ the network weights. 
After substituting the DIP model (\ref{DIP}) into  (\ref{PET model}), the PET forward projection model becomes
\beq
\overline{\y} = \Pm \bet(\tht|\z) + \rv.
\eeq

The maximum-likelihood estimate of the unknown $\tht$ is obtained by
\beq
\hat{\tht} = \arg\max\limits_{\tht}L\Big(\y| \bet(\tht|\z)\Big).
\eeq
Once $\tht$ is estimated, the PET activity image is calculated as
\beq
\hat{\x} = \bet(\hat{\tht}|\z). 
\eeq

\begin{figure}[t]
	\centering
	\includegraphics[trim=1cm 0cm 1cm 0cm, width=2.2in]{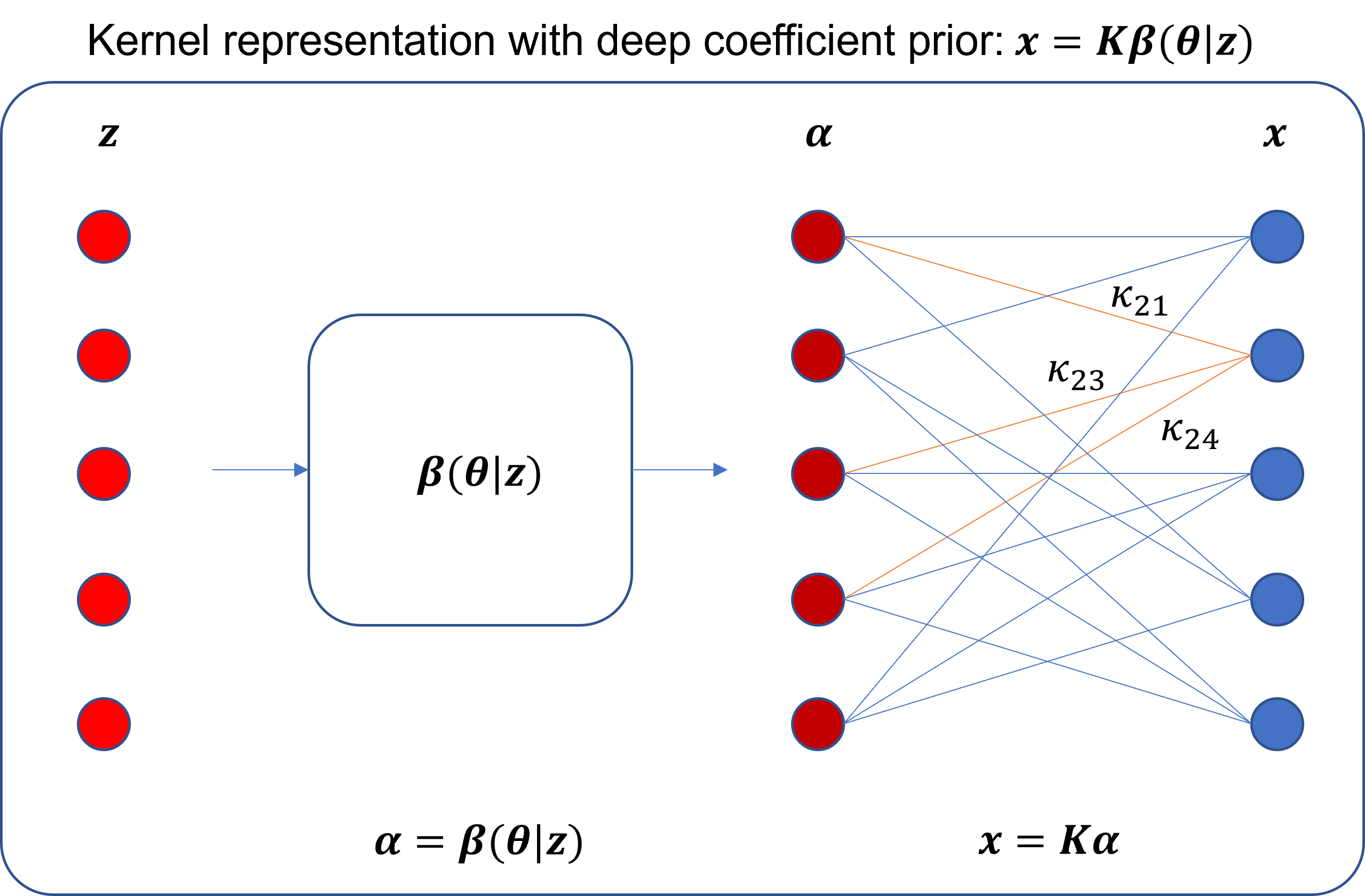}
	\caption{Graphical illustration of the kernel representation with deep coefficient prior.}
	\label{fig:1}	
\end{figure}

To solve the resulting nonlinear optimization problem,  Gong {\em et al.} \cite{Gong2019} use the ADMM algorithm,
\beq
\x^{n+1}= \arg\max\limits_{\x}L(\y|\x) - \frac{\rho}{2}||\x-\bet(\tht^n|\z)+\muv||^2, 
\label{sub_p}
\eeq
\beq
\tht^{n+1}=\arg\min\limits_{\tht}||\bet(\tht|\z)-(\x^{n+1}+\muv^n)||^2,
\label{sub_d}
\eeq
\beq
\muv^{n+1} = \muv^n +\x^{n+1} -\bet(\tht^{n+1}|\z),
\eeq
where the subproblem (\ref{sub_p}) is a penalized-likelihood image reconstruction problem and the subproblem (\ref{sub_d}) is an image-domain DIP learning using a mean-square error (MSE) loss function. $\rho$ is a hyper-parameter. One well-known weakness of the ADMM algorithm is that $\rho$ is usually difficult to tune.

\section{Proposed Neural KEM}
\subsection{Kernel Method with Deep Coefficient Prior}

In this work, we propose to describe the kernel coefficient image $\alp$ in the kernel method as a function of neural networks,
\beq
\alp = \bet(\tht|\z),
\eeq
where $\z$ and $\tht$ are again the input (e.g., the composite image prior in dynamic PET \cite{Wang2015}) and weights of the neural network $\bet$, sharing the same spirit of the DIP model \cite{Gong2019}. 
This provides a kernel representation with deep coefficient prior for PET image,
\beq
\x = \K\bet(\tht|\z).
\label{dcp}
\eeq
Fig. \ref{fig:1} shows a graphical illustration of the proposed model using neural network layers, of which the last layer is linear and has fixed network weights as determined by the kernel matrix $\K$. 

The proposed model (\ref{dcp}) becomes the DIP model in \cite{Gong2019} if the kernel matrix $\K$ is an identity matrix; the model is also equivalent to the standard kernel model \cite{Wang2015} if the neural network $\bet$ is an identity mapping. 
When a more complex neural network model (e.g., U-net) is used, the deep coefficient prior then introduces an implicit regularization to stabilize the estimation of the kernel coefficient image $\alp$. 

By substituting the proposed image model in (\ref{dcp}) into the standard PET forward projection model in (\ref{PET model}), we obtain the following forward model for PET image reconstruction,
\beq
\overline{\y} = \Pm\K\bet(\tht|\z) + \rv.
\label{dcp_model}
\eeq
The unknown $\tht$ of the neural network is estimated from the projection data by maximizing the Poisson log-likelihood, 
\beq
\hat{\tht} = \arg\max\limits_{\tht}L\Big(\y|\K \bet(\tht|\z)\Big).
\label{eq-lik}
\eeq
Once $\hat{\tht}$ is estimated, the PET image is obtained by 
\beq
\hat{\x} = \K \bet(\hat{\tht}|\z).
\eeq 

\subsection{Tomographic Reconstruction of Neural Networks Using Optimization Transfer}
The optimization problem in (\ref{eq-lik}) is challenging to solve because the unknown $\tht$ is non-linearly involved in the projection domain. One possible solution is the ADMM algorithm as used in \cite{Gong2019} but tuning the hyperparameter $\rho$ is nontrivial in practice. Here we develop an easy-to-implement optimization transfer algorithm using a similar concept as used for nonlinear parametric PET image reconstruction of tracer kinetics \cite{Wang2008, Wang2012} and for joint image registration and reconstruction \cite{Fessler2010}.

\begin{figure}[t]
	\centering
	\includegraphics[trim=2cm 0cm 2cm 0cm, width=1.2in]{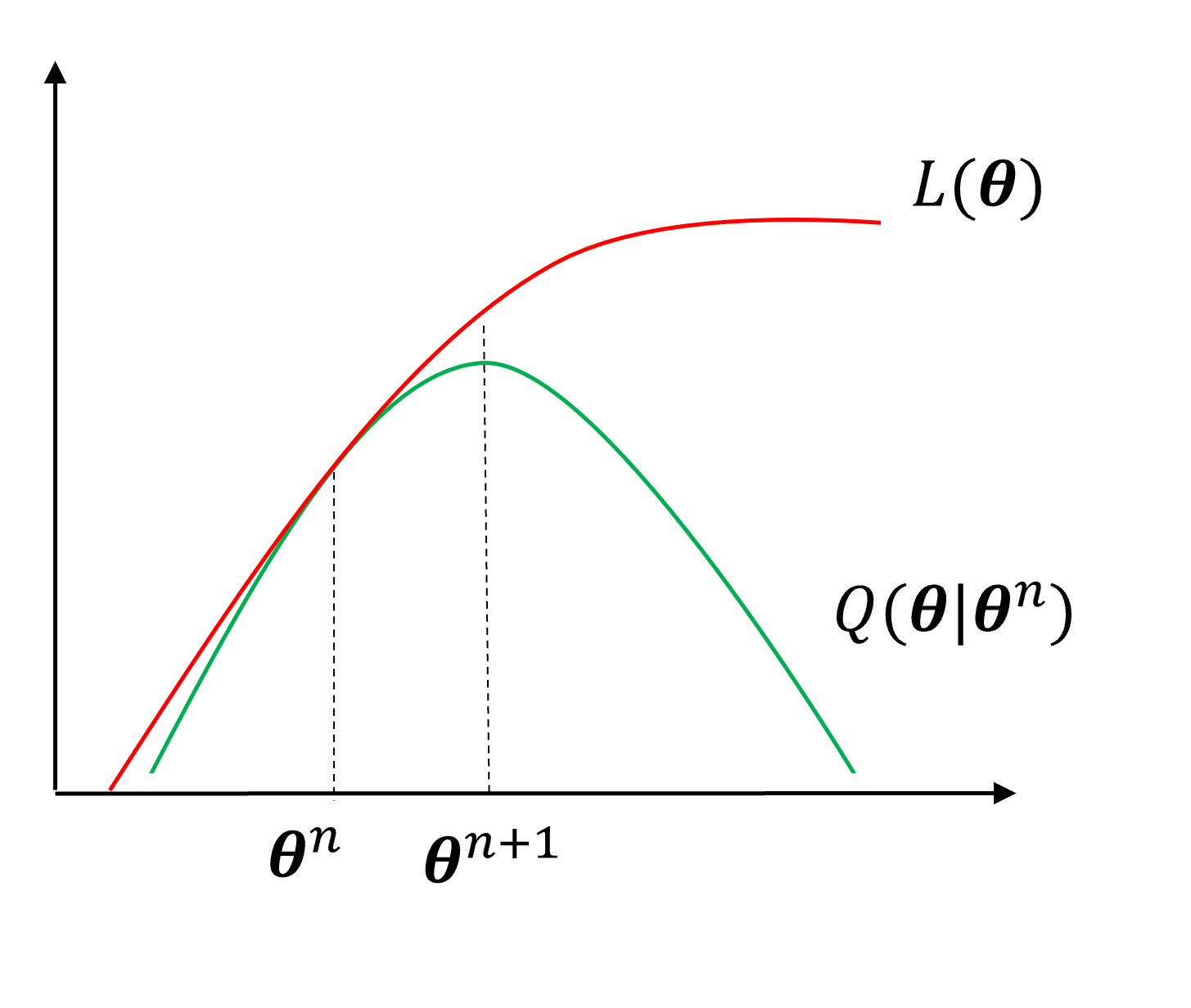}
	\vspace{-15pt}	
	\caption{\txtb{Illustration of optimization transfer used in this work. The surrogate function $Q(\tht|\tht^n)$ minorizes the original likelihood function $L(\tht)$. $Q(\tht|\tht^n)$ is designed to be easier to optimize. The solution $\tht^{n+1}$ guarantees a monotonic increase in $L$.}}
	\label{OT}

\end{figure}

\txtb{The basic idea of optimization transfer \cite{Lange2000} is to construct a surrogate function $Q(\tht|\tht^n)$ at iteration $n$, which minorizes the original objective function $L(\tht)$, as illustrated in Fig. \ref{OT}.}  Following the concave property of the log function \cite{Lange2000} (also see Eq. 9 in \cite{Wang2012}) and treating $\Pm\K$ as a single matrix $\A$ with its ($i,j$)th element being $a_{ij}$, we have the following inequality, 
\beq
\begin{aligned}
	\log\overline{y}_i&=\log\Big(\sum_{j=1}^J a_{ij}\beta_j(\tht|\z) + r_i\Big)\\
	&\geq\sum_{j=1}^J\frac{a_{ij}\beta_j(\tht^n|\z)}{\overline{y}_i^n}\log\beta_j(\tht|\z) + c_i^{n},
\end{aligned}
\label{scaled}
\eeq
where $\overline{y}_i^n = [\A\bet(\tht^n|\z) + \rv]_i$. 
The iteration-dependent constant,
\beq
c_i^n=\Big(\sum_{j=1}^J\frac{a_{ij}\beta_j(\tht^n|\z)}{\overline{y}_i^n}\log\frac{\overline{y}_i^n}{\beta_j(\tht^n|\z)}\Big) + \frac{r_i}{\overline{y}_i^n} \log \overline{y}_i^n,
\eeq
is independent of the unknown parameter $\tht$ and is thus omitted hereafter.

Based on (\ref{scaled}), an EM-type surrogate function $Q(\tht|\tht^n)$ can be built for the original likelihood function $L$ \txtb{in a similar way as used in \cite{Wang2012}},
\beq
\txtc{Q(\tht|\tht^n) = \sum_{j=1}^J w_j\Big(\hat{\alpha}_j^{n+1} \log \beta_j(\tht|\z) - \beta_j(\tht|\z)\Big)},
\label{surrogate}
\eeq
where $w_j$ corresponds to the $j$th pixel of $\w$ defined in (\ref{eq-w}). $\hat{\alp}^{n+1} $ is an intermediate kernel coefficient image updated with
\beq
\hat{\alp}^{n+1} = \frac{\alp^n}{\w}\cdot\left( \K^T\Pm^T\frac{\y}{\overline{\y}^n}\right),
\label{eq-KEM}
\eeq
which is one iteration of KEM defined by (\ref{KEM}) with $\alp^n \triangleq \bet(\tht^n|\z)$.

The surrogate $Q(\tht|\tht^n)$ resembles an image-domain Poisson log-likelihood function (with a pixel-wise weight $\w$). Using (\ref{scaled}), it is straightforward to prove that the surrogate satisfies the following two conditions for optimization transfer \cite{Lange2000},
\beq
\txtb{Q(\tht|\tht^n) - Q(\tht^n|\tht^n) \leq L\big(\y|\K \bet(\tht|\z)\big) - L\big(\y|\K \bet(\tht^n|\z)\big),}
\label{ot}
\eeq
\beq
\nabla Q(\tht^n|\tht^n) = \nabla L\big(\y|\K \bet(\tht^n|\z)\big),
\eeq
\txtb{where $\nabla$ denotes the gradient with respect to $\tht$.}

The original optimization problem in (\ref{eq-lik}) is now equivalently transferred into the maximization of the surrogate function (\ref{surrogate}) at each iteration $n$,
\beq
\tht^{n+1} = \arg\max\limits_{\tht}Q(\tht|\tht^n),
\label{op_problem}
\eeq
which performs an image-domain neural-network learning for seeking a $\bet$ approximation to the intermediate kernel coefficient image $\hat{\alp}^{n+1}$. \txtb{The learning can be implemented using any existing optimization algorithm (e.g., the Adams optimizer) that is available in a deep-learning software library such as PyTorch or TensorFlow.} \txtb{Because of $Q(\tht^{n+1}|\tht^n)\geq Q(\tht^n|\tht^n)$ and (\ref{ot}),} 
the surrogate optimization guarantees \txtb{convergence to a local optimum} and a monotonic increase in the original likelihood $L$,
\beq
L\Big(\y|\K \bet(\tht^{n+1}|\z)\Big) \geq L\Big(\y|\K \bet(\tht^{n}|\z)\Big).
\eeq

\begin{figure}[t]
	\centering
	\includegraphics[trim=2cm 0cm 2cm 0cm, width=2.5in]{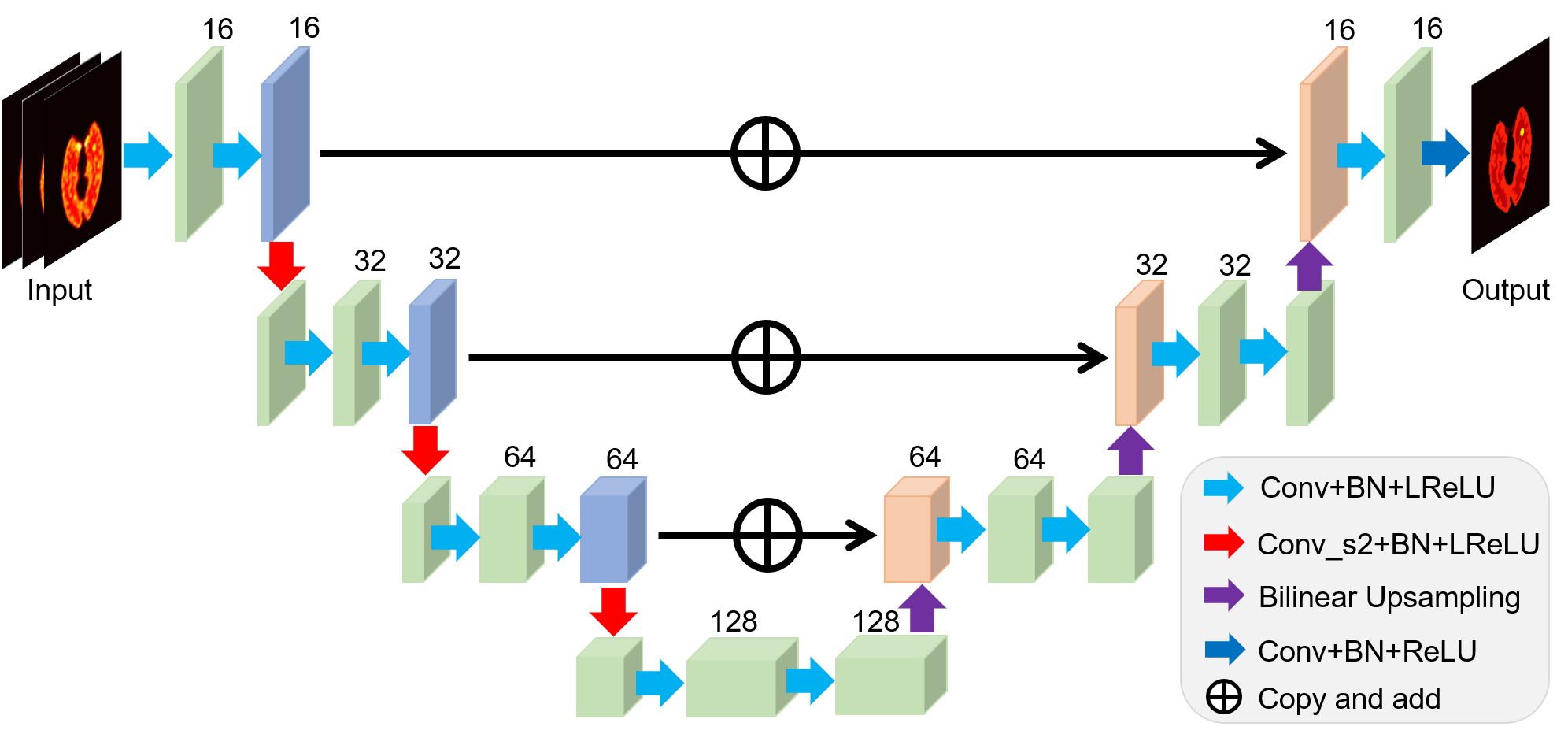}
	\caption{Illustration of the modified residual U-net $\bet(\tht|\z)$ used in this work.}
	\label{fig:2}	
\end{figure}
\subsection{Summary of the Algorithm and Implementations}

A pseudo-code of the proposed algorithm is provided in Algorithm 1. Each iteration of the algorithm consists of two separate steps:
\begin{enumerate}
	\item Image reconstruction: Obtain an intermediate kernel coefficient image update $\hat{\alp}^{n+1}$ from the projection data $\y$ using KEM in (\ref{eq-KEM});
	\item Neural-network learning: Find a CNN approximation of  the intermediate image $\hat{\alp}^{n+1}$ using the image-domain maximum-likelihood optimization in (\ref{op_problem}).
\end{enumerate}
We call this algorithm {\em Neural KEM} to reflect the fact that neural-network learning is used following the KEM update. Compared to ADMM, the Neural KEM  algorithm does not need to tune a hyperparameter and is easier to use.

\txtb{The proposed algorithm is applicable to different neural network
architectures that are suitable for image representation.} In our work, a \txtb{popular} residual U-net (e.g., used in \cite{Gong2019}) is used for neural network learning and is illustrated in Fig. \ref{fig:2}. The network is available in both 2D and 3D versions for learning 2D and 3D images, respectively. It consists of the following operations: 1) 3$\times$3 ($\times$3) 2D (3D) convolutional layer, 2) 2D (3D) batch normalization (BN) layer, 3) leaky rectified linear unit (LReLU) layer, 4) 3$\times$3 ($\times$3) convolutional layer with stride 2$\times$2 ($\times$2) for down-sampling, 5) 2$\times$2 ($\times$2) bilinear (trilinear) interpolation layer for up-sampling, 6) identity mapping layer that adds feature maps from left-side encoder path to the right-side decoder path. In addition, a ReLU layer is used before the output in order to satisfy the non-negative constraint on the kernel coefficient image. 
\txtb{The total number of model parameters in the 3D U-net is about 1.3 million.}

\begin{table}[t]
	\normalsize
	\begin{tabularx}{3.4in}{p{0.1cm}X}
		\toprule
		&{\bf Algorithm 1} Neural KEM for PET reconstruction\\
		\midrule
		1: & Input parameters: Maximum iteration number {\ttfamily MaxIt}, initial $\tht^1$.\\
		2: & {\bf for} $n = 1$ to {\ttfamily MaxIt} {\bf do}\\
		3: & ~~ Obtain an intermediate coefficient image update:\\
		& ~~ $\hat{\alp}^{n+1} = \frac{\alp^n}{\w}\cdot\left( \K^T\Pm^T\frac{\y}{\Pm\K\alp^n+\rv}\right)$,\\
		& ~~ with $\alp^{n} = \bet(\tht^{n}|\z)$ \\
		4: & ~~ \txtc{Perform a neural network learning by maximizing:}\\
		& ~~ \txtc{$Q(\tht|\tht^n)\triangleq\sum_{j} w_j\Big(\hat{\alpha}_j^{n+1} \log \beta_j(\tht|\z) - \beta_j(\tht|\z)\Big)$}\\
		5: & {\bf end for}\\
		6: & {\bf return} $\hat{\x} = \K \bet(\hat{\tht}|\z)$ \\
		\bottomrule
	\end{tabularx}
\end{table}

\begin{figure}[t]
	\vspace{-0pt}
	\centering
	\subfloat[]{\includegraphics[trim=1.2cm 0cm 0cm 0cm, width=1.7in]{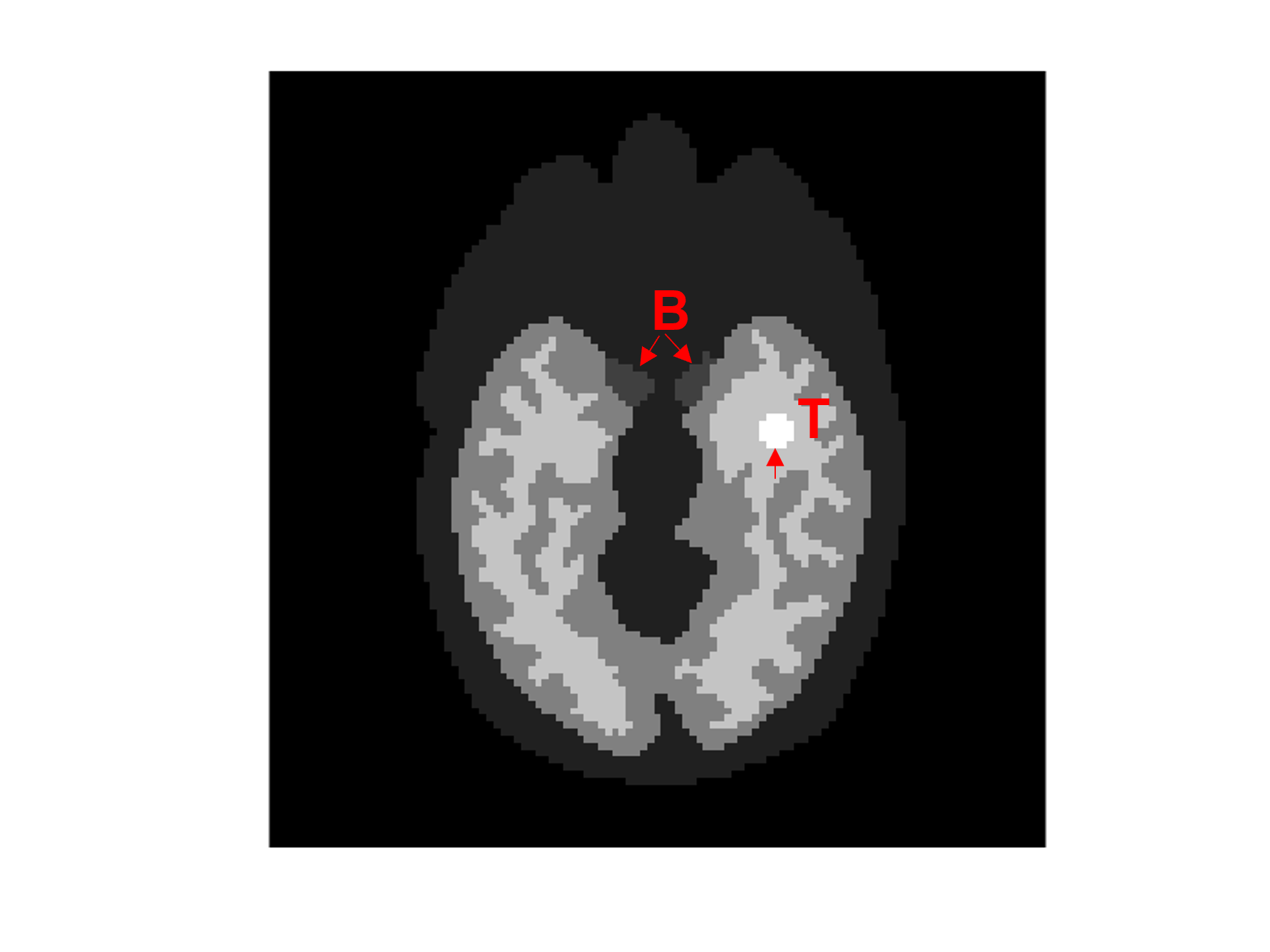}}
	\subfloat[]{\includegraphics[trim=1cm 0cm 1cm 0cm, width=1.6in]{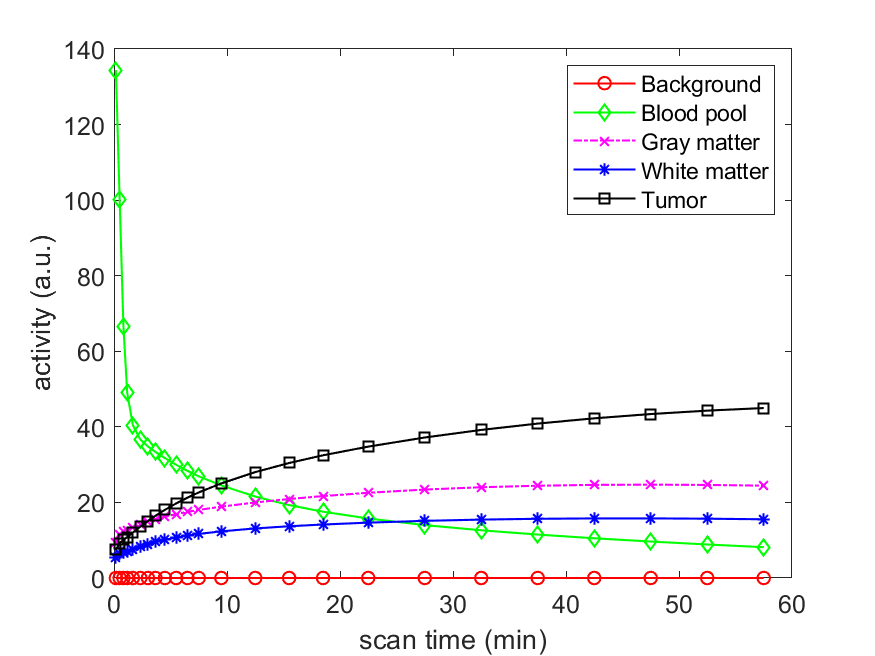}}
	\caption{Digital phantom and time activity curves used in the simulation studies. (a) Zubal brain phantom; \txtb{`B' represents the blood ROI and `T' is the tumor ROI.} (b) Regional time activity curves.}
	\label{fig:phant}	
\end{figure}

\begin{figure*}[htp]
	\vspace{-0pt}
	\centering
	\includegraphics[trim=0cm 0cm 0cm 0cm, 
		clip, width=7in]{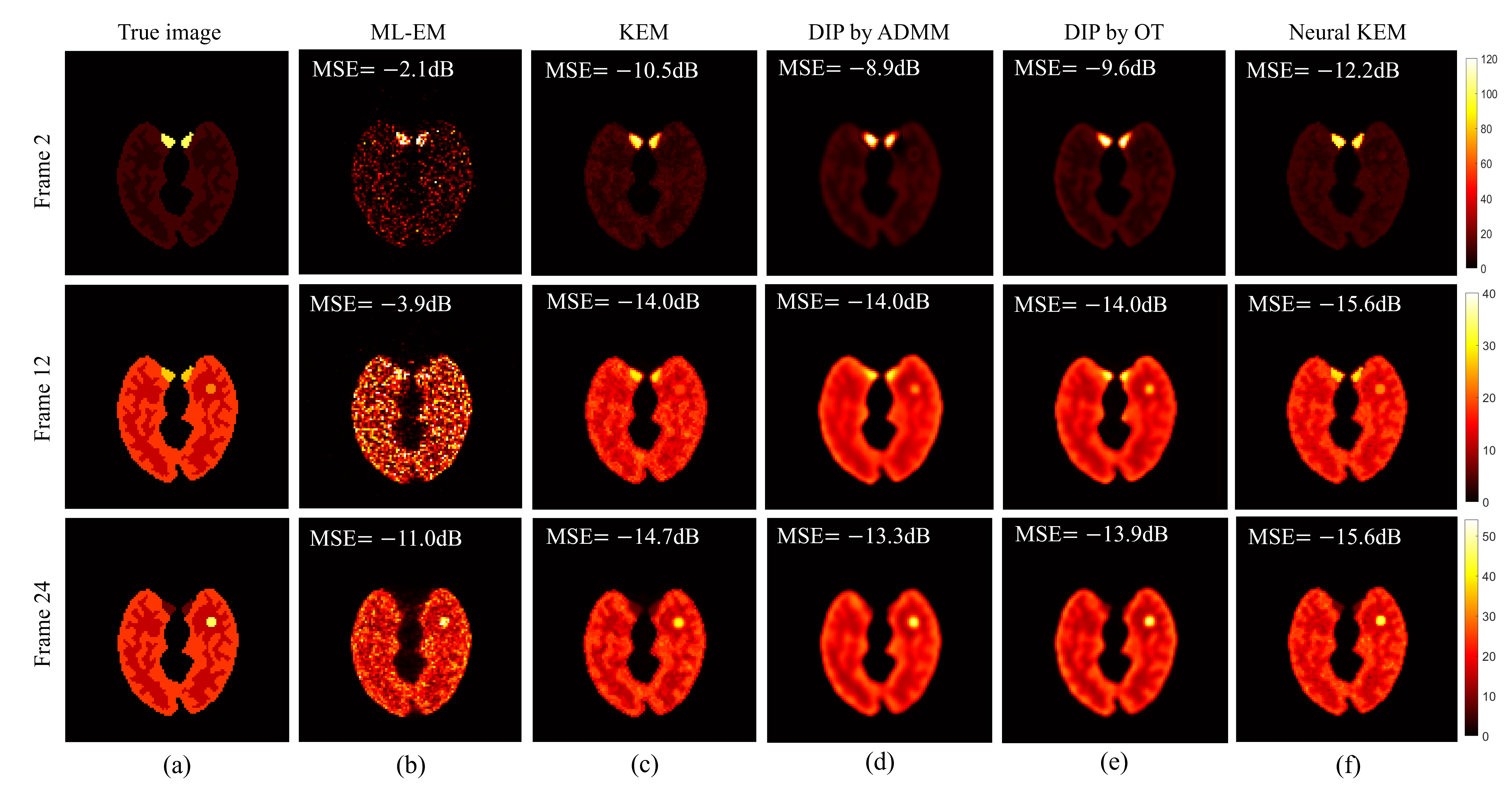}
		\label{fig_1_case}
	\caption{True activity images and reconstructed images by different reconstruction methods for frame 2 (top row), frame 12 (middle row) and frame 24 (bottom row). (a) True images, (b) ML-EM, (c) KEM \cite{Wang2015}, (d) DIP by ADMM \cite{Gong2019}, (e) DIP by OT, (f) Proposed neural KEM.}
	\label{fig:4}
\end{figure*}

\section{Validation Using \txtb{2D} Computer Simulation}
\subsection{Simulation Setup}
We conducted a \txtb{two-dimensional (2D)} computer simulation study to validate the proposed method in dynamic PET image reconstruction. Dynamic scans were simulated for a GE DST whole-body PET scanner using a Zubal head phantom shown in Fig. \ref{fig:phant}a. \txtb{The phantom is
composed of gray matter, white matter, blood pools (18mm in long axis) and a tumor (15 mm in diameter). The detector system consists of 280 detector blocks, arranged as four rings of 70 blocks each. The width of the block is 38.35mm. The scanner consists of 10,080 BGO crystals.} A one-hour dynamic scan was divided into 24 time frames: 4$\times$20s, 4$\times$40s, 4$\times$60s, 4$\times$180s, and 8$\times$300s. The pixel size is 3$\times$3 mm$^2$ and the image size is 111$\times$111. The time activity curve of $^{18}$F-FDG in each region is shown in Fig. \ref{fig:phant}b. Dynamic activity images were first forward projected to generate noise-free sinograms. Poisson noise was then introduced. Scatters were simulated using the SimSET package\cite{Harrison1993} \txtb{using the cylindrical scanner model with no block and gap effects included}. We also included 20\% uniform random events. 
Attenuation map, mean scatters and randoms were used in all reconstruction methods to obtain quantitative images. The expected total number of events over 60 min was 8 million. Ten noisy realizations were simulated and each was reconstructed independently for comparison.

\subsection{Reconstruction Methods}

We compared the proposed neural KEM with four different reconstruction methods: (1) standard ML-EM, (2) KEM \cite{Wang2015}, (3) DIP reconstruction by ADMM \cite{Gong2019}, and (4) DIP reconstruction by the optimization transfer (OT) algorithm, which is equivalent to the neural KEM with $\K = \I$.

In the kernel-based methods (regular KEM and neural KEM), three 20-minute composite frames were used to generate the image prior data $\z$ as used in \cite{Wang2015}. \txtb{The radial Gaussian kernel function $\kappa(\f_j, \f_l)=  {\rm exp}(-||\f_j - \f_l||^2/2\sigma^2)$ was used.} Pixel intensity values extracted from the composite images were used to form the feature vector $\f$ for generating the kernel matrix $\K$ using \txtb{$\sigma = 1$ and kNN with $k$=48 which were the same as used in \cite{Wang2015}. }

For the DIP reconstruction by ADMM \cite{Gong2019}, within each outer iteration, 4 iterations were used for solving  (\ref{sub_p}) and 50 iterations were used for solving (\ref{sub_d}). These settings were empirically optimized for obtaining stable results according to image mean squared error (MSE; defined in next subsection) in our experiments. The effect of the ADMM hyper-parameter $\rho$ was also investigated and reported. $\rho=0.05$ was chosen for nearly optimal image MSE.

The input of CNN in both the DIP methods and neural KEM was set to the composite image prior $\z$. A ML-EM image was also tested but resulted in worse results. For implementation, the tomographic reconstruction step was implemented in MATLAB and the neural-network learning step was implemented in PyTorch, both on a PC with an Intel i9-9920X CPU with 64GB RAM and a NVIDIA GeForce RTX 2080Ti GPU. The Adam algorithm was used with a learning rate $10^{-3}$ for neural network learning.  All reconstructions were run for 60 iterations with a uniform initial image. The subiteration number for the neural network learning step was empirically optimized to be 150 in both the DIP by OT and neural KEM method for image MSE. The effect of this subiteration number was also investigated. 


\begin{figure*}[t]
	\vspace{-0pt}
	\centering
	\subfloat[]{\includegraphics[trim=0.5cm 0cm 1cm 0.5cm, clip,height=1.5in]{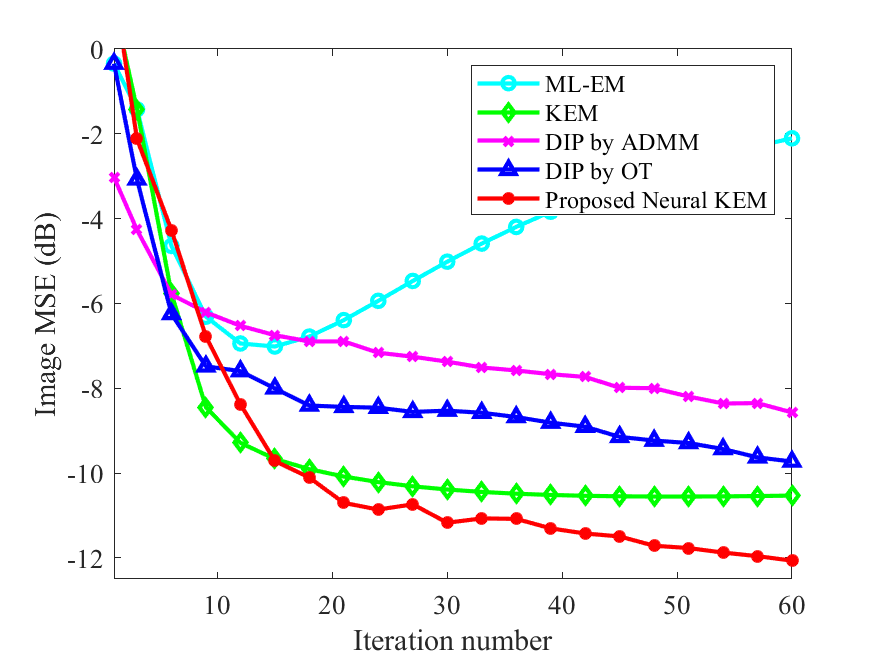}
		\label{fig_3_case}}
	\hfil
	\subfloat[]{\includegraphics[trim=0.5cm 0cm 1cm 0.5cm, 
		clip, height=1.5in]{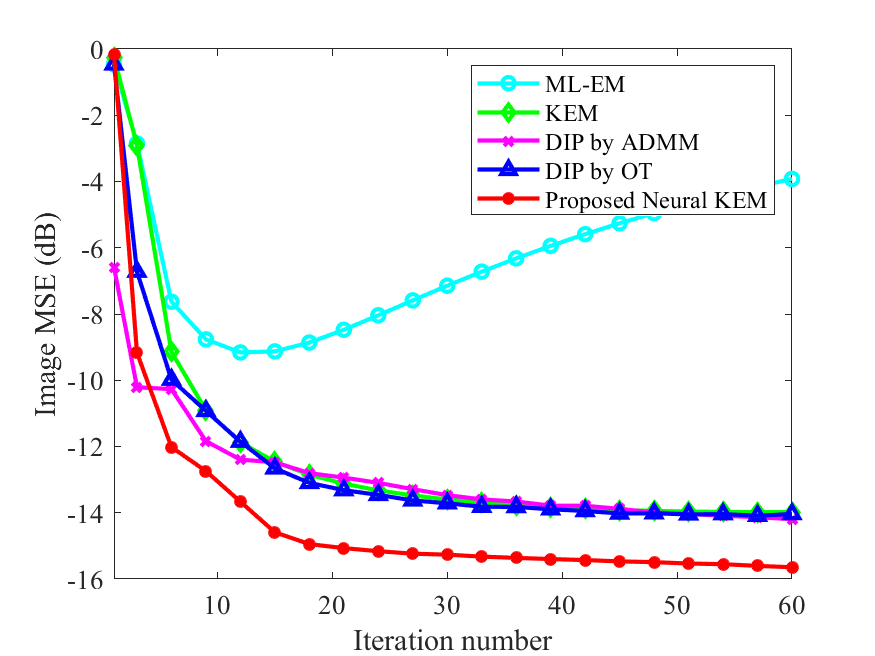}
		\label{fig_5_case}}
	\hfil
	\subfloat[]{\includegraphics[trim=0.5cm 0cm 1cm 0.5cm, height=1.5in]{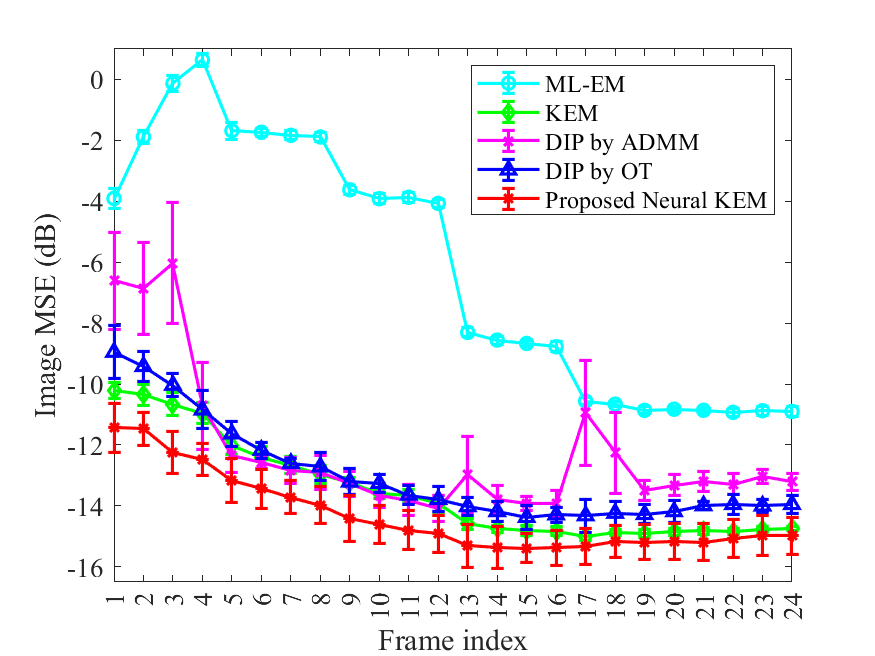}}
	\caption{Plots of image MSE for different reconstruction methods. (a-b) image MSE as a function of iteration number for frame 2 (a) and frame 12 (b); (c) image MSE of all time frames. The error bars in (c) were obtained from 10 realizations.}
	\label{fig:5}
\end{figure*}
\begin{figure*}[t]
	\vspace{-0pt}
	\centering
	\subfloat[Blood ROI quantification]{\includegraphics[trim=0.5cm 0cm 0.5cm 0cm, clip, width=6.5in]{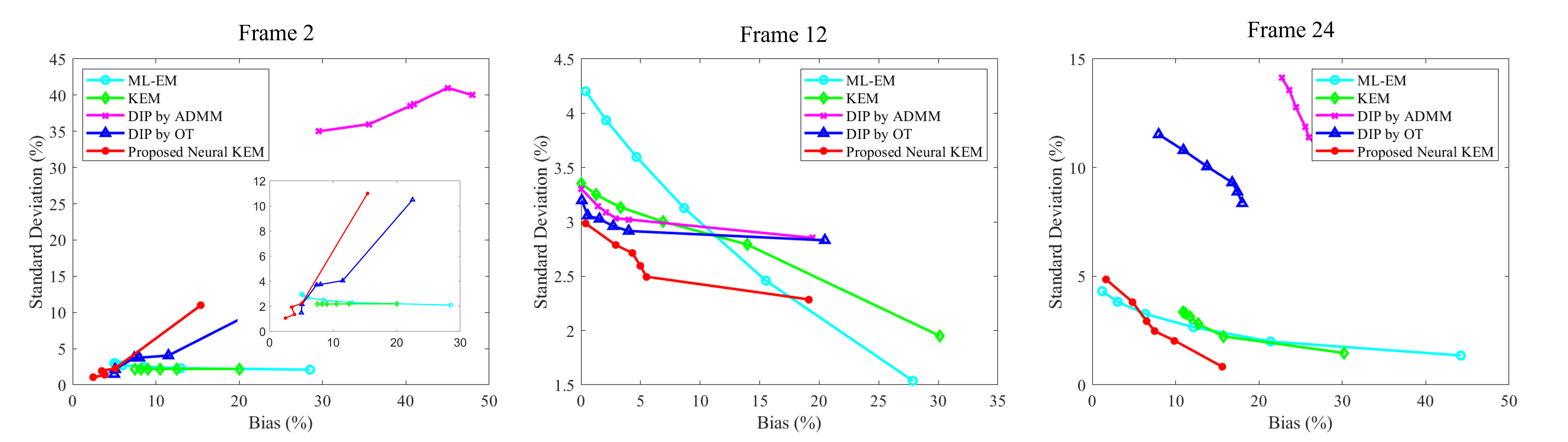}
		\label{fig_4_case}}\\
	\subfloat[Tumor ROI quantification]{\includegraphics[trim=0.5cm 0cm 0.5cm 0cm, 
		clip, width=6.5in]{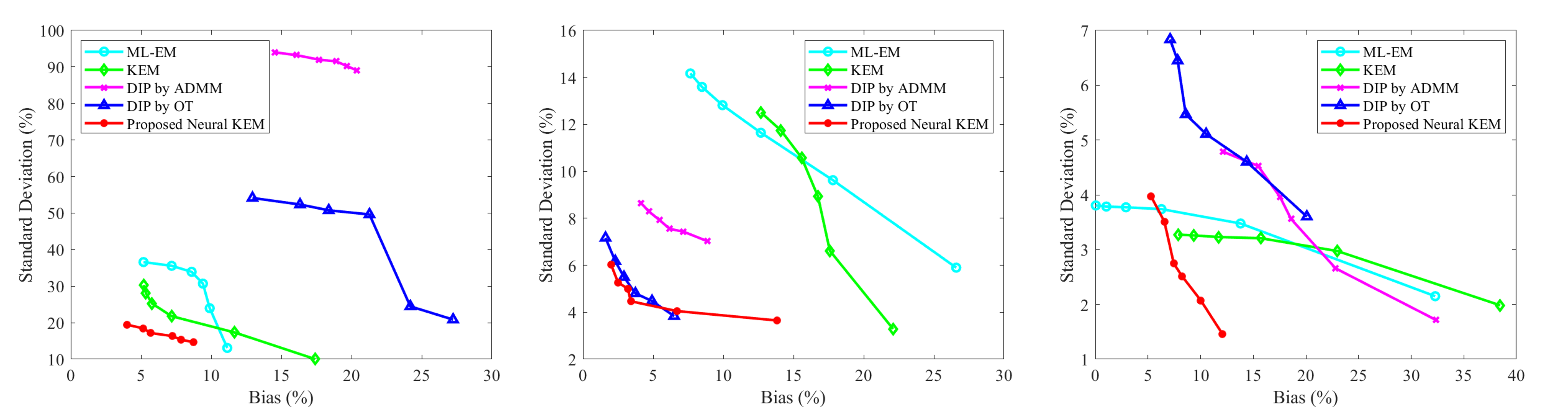}
		\label{fig_4_case}}
	\caption{\txtb{Plots of bias-SD trade-off for ROI quantification in frame 2, 12 and 24 by varying the iteration number from 10 to 60 (i.e., from rightmost to leftmost on each curve). (a) Blood ROI quantification, (b) tumor ROI quantification. A zoom-in is included for frame 2 in (a).}}
	\label{fig:7}
\end{figure*} 

\subsection{Evaluation Metrics}
Different methods were first compared using image MSE defined by
\beq
{\rm MSE}(\hat{\x}_m) = 10\log_{10}\big( ||\hat{\x}_m - \x_m^{\rm{true}}||^2/||\x_m^{\rm{true}}||^2\big) (\mathrm{dB}),
\eeq
where $\hat{\x}_m$ is an image estimate of frame $m$ obtained with one of the reconstruction methods and $\x_m^{\rm{true}}$ denotes the ground truth image.
The ensemble bias and standard deviation (SD) of the mean intensity in regions of
interest (ROIs) were also calculated to evaluate ROI quantification,
\beq
\rm{Bias} = \frac{1}{\rm{c}^{\rm{true}}}\big|\overline{\rm{c}} - \rm{c}^{\rm{true}}\big|,\;
{\rm SD} =\frac{1}{\rm{c}^{\rm{true}}}\sqrt{\frac{1}{N_r-1}\sum_{i=1}^{N_r}|\rm{c}_{\it{i}}-\overline{\rm{c}}|},
\eeq
where $\rm{c}^{\rm{true}}$ is the noise-free intensity and $\overline{\rm{c}}=\frac{1}{N_r}\sum_{i=1}^{N_r}\rm{c}_{\it{i}}$ denotes the mean of $N_r$ realizations. $c_i$ is the mean ROI uptake in the $i$th realization and $N_r = 10$ in this study.

\subsection{Comparison for Reconstructed Image Quality}

Fig. \ref{fig:4} shows the true activity images and reconstructed images at iteration 60 by five different reconstruction methods for frame 2 (early 20-s frame, low count level), frame 12 (middle 1-min frame, moderate count level) and frame 24 (late 3-min frame, relatively high count level), respectively. The results of MSE in dB are included. As expected, kernel-based methods (regular KEM and neural KEM) achieved a better image quality with lower MSE as compared to the methods without kernel ((b), (d) and (e)). The DIP by ADMM \cite{Gong2019} and DIP by OT both suppressed noise well but also led to over-smoothness. The proposed neural KEM was less noisy than the regular KEM due to the added level of regularization from the deep coefficient prior on $\alp$ and demonstrated better detail preservation than the DIP methods due to the additional structural information embedded in the kernel matrix $\K$.

Fig. \ref{fig:5}(a) and Fig. \ref{fig:5}(b) further show image MSE as a function of iteration number for the two different frames (frame 2 and frame 12). 
For the DIP reconstruction, the ADMM algorithm demonstrated a relatively faster convergence rate in early iterations than the KEM and DIP by OT. This is mainly because four sub-iterations were used for the tomographic reconstruction step in the ADMM algorithm while one sub-iteration was used for other algorithms. The DIP by OT was either comparable to (Fig. \ref{fig:5}(b)) or better than (Fig. \ref{fig:5}(a)) the DIP by ADMM. Note that here the DIP by ADMM and DIP by OT were not always close to each other. This can be explained by that the neural network model is nonlinear and different algorithms are not guaranteed to provide the same solution.

Fig. \ref{fig:5}(c) shows the plots of image MSE for all time frames reconstructed by different methods with 60 iterations. Error bars were calculated over 10 noisy realizations. The DIP by ADMM showed a slightly unstable behavior across different frames. This is likely because a single value of the hyper-parameter $\rho$ has varying efficacy for different time frames. The regular KEM was either equivalent to or slightly better than the DIP by OT. The proposed neural KEM further improved all the frames as compared to the regular KEM.

\subsection{Comparison for ROI Quantification}

Fig. \ref{fig:7} shows the trade-off between the bias and SD of different methods for ROI quantification in \txtb{the blood and tumor regions. The two ROIs are the same as the anatomical regions marked in Fig. \ref{fig:phant}a.} The curves were obtained by varying the iteration number from 10 to 60 iterations with an interval of 10 iterations. As expected, the bias decreases as the iteration number increases in all the methods. For the blood ROI quantification, the DIP by OT was better than the DIP by ADMM because of the improved convergence and stability by OT in these \txtb{three cases.} For the tumor ROI quantification, the DIP by OT was either better than the DIP by ADMM in the low and medium cases (frame 2 and frame 12) or comparable in the high-count case (frame 24).

\txtb{The proposed neural KEM achieved the lowest SD and bias simultaneously for frame 2 and frame 12 after 20 iterations, demonstrating its advantage for low- and medium-count frames. In the high-count case (frame 24), the traditional ML-EM achieved the lowest bias for small targets (i.e., the tumor and blood ROIs) due to a good recovery of the contrast at a high iteration number. The neural KEM was with a higher bias due to oversmoothness in this high-count case but it was still better than the regular KEM and DIP methods.}

\subsection{Effect of Method Parameters}

\begin{figure}[t]
	\vspace{-10pt}
	\centering
	\subfloat[]{\includegraphics[trim=0.5cm 0cm 1cm 0cm, clip,width=1.6in]{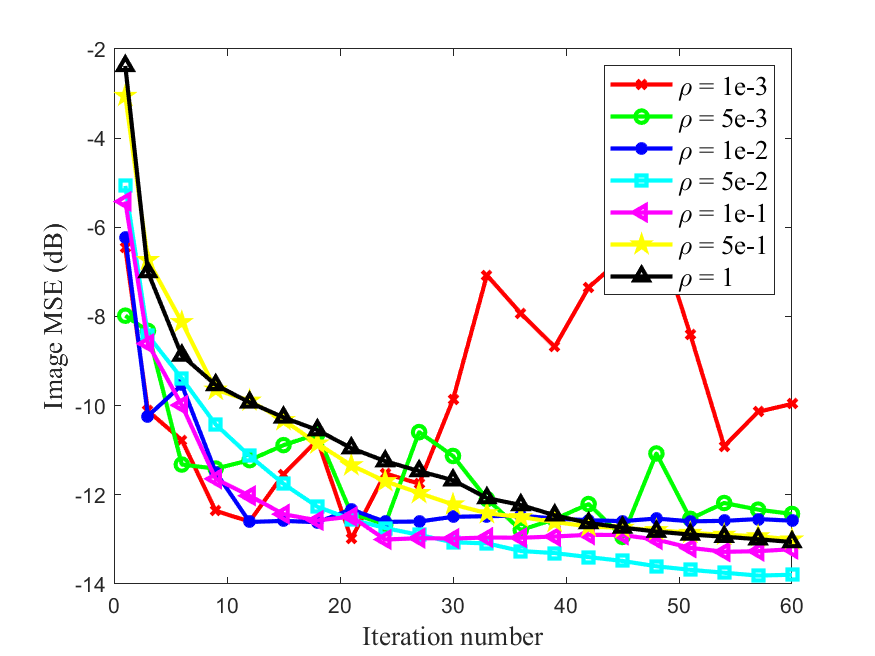}
		\label{fig_3_case}}
	\hfil
	\subfloat[]{\includegraphics[trim=0.5cm 0cm 1cm 0cm, 
		clip, width=1.6in]{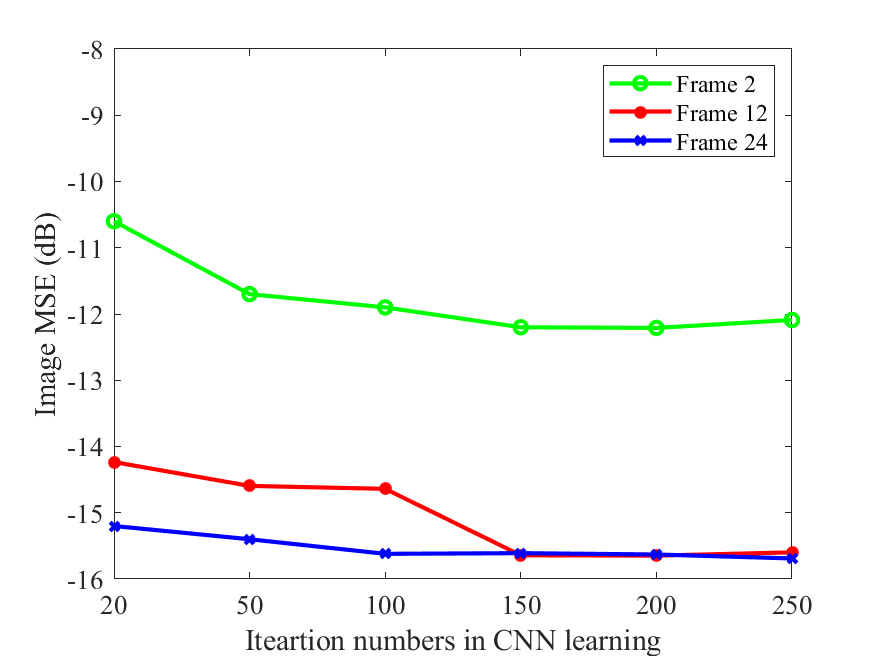}
		\label{fig_4_case}}
	\caption{Effect of (a) ADMM hyper-parameter $\rho$ of the DIP method and \txtb{(b) CNN learning subiterations of the neural KEM on the image MSE.}}
	\label{parameter}
\end{figure}

To demonstrate the challenge for choosing a proper ADMM parameter $\rho$ in the DIP method, Fig. \ref{parameter}(a) shows the iteration-based MSE result for a range of $\rho$ values for frame 12. An inappropriate $\rho$ resulted in instability and oscillations across iterations. $\rho = 0.05$ was a good choice for this frame but resulted in poor MSE for some of other frames, as shown in Fig. \ref{fig:5}(c). The OT algorithm avoids this difficult-to-tune parameter.

Similar to other methods that use neural network-based deep prior, the proposed neural KEM involves a sub-iteration number that needs to be determined for the neural-network learning step. Fig. \ref{parameter}(b) shows the effect of this sub-iteration number on the final image MSE for \txtb{frame 2 (low-count), frame 12 (medium-count) and frame 24 (high-count).} The result suggests a reasonable choice is 150 iterations, which also worked well for other frames. We also found that the CNN training is stable when the learning rate in the Adam optimizer ranges from 10$^{-4}$ to 10$^{-2}$. 

\begin{figure*}[htp]

	\centering
	\includegraphics[trim=0cm 0cm 0cm 0cm, clip,width=6.0in]{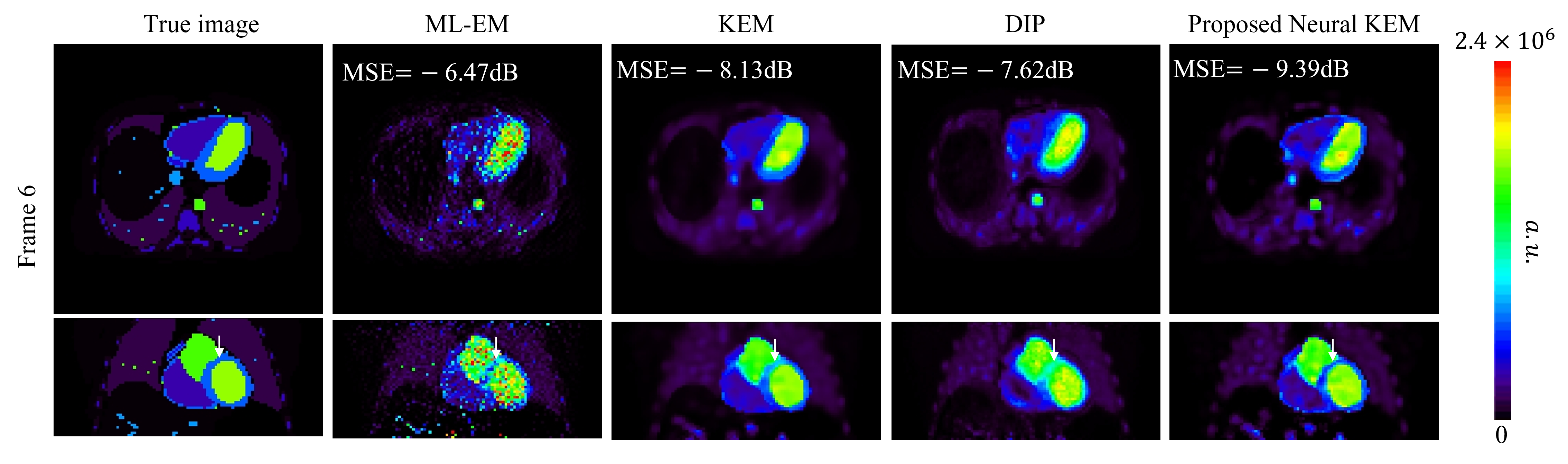}
		\label{fig_1_case}\\
	\includegraphics[trim=0cm 0cm 0cm 0cm, 
		clip, width=6.0in]{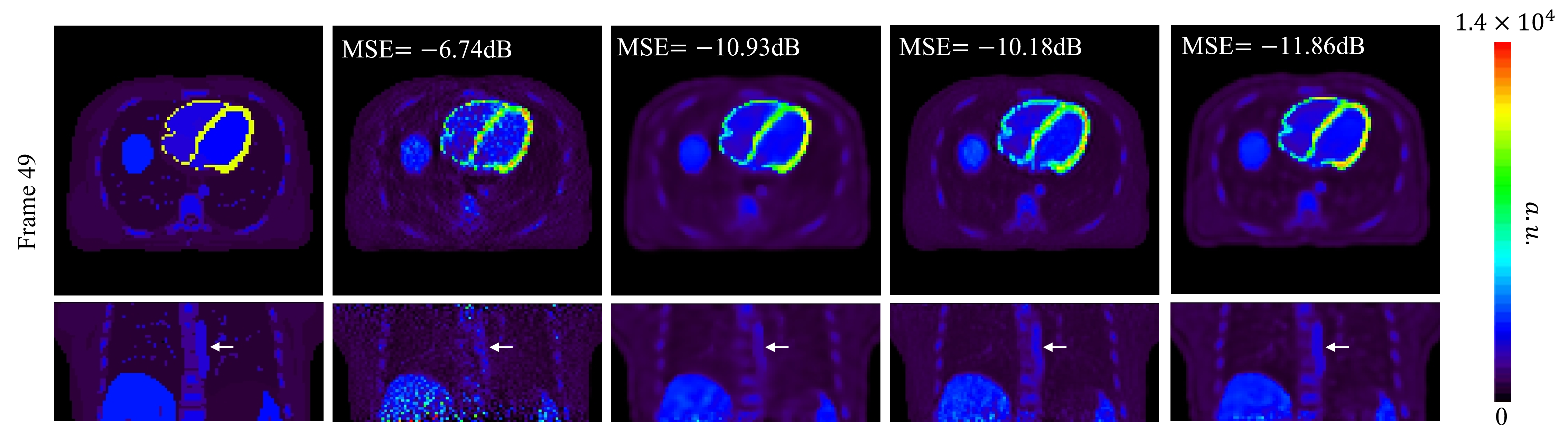}
		\label{fig_2_case}
	\caption{\txtb{True activity images and reconstructed images by different methods in a 3D computer simulation study.  A low-count case (frame 6, top row) and a high-count case (frame 49, bottom row) are shown. Each image is shown in transverse and coronal views.}}
	\label{3D simulation}
\end{figure*}

\begin{figure*}[h]
	\centering
	\subfloat[]{\includegraphics[trim=1cm 0cm 1cm 0cm, height=1.7in]{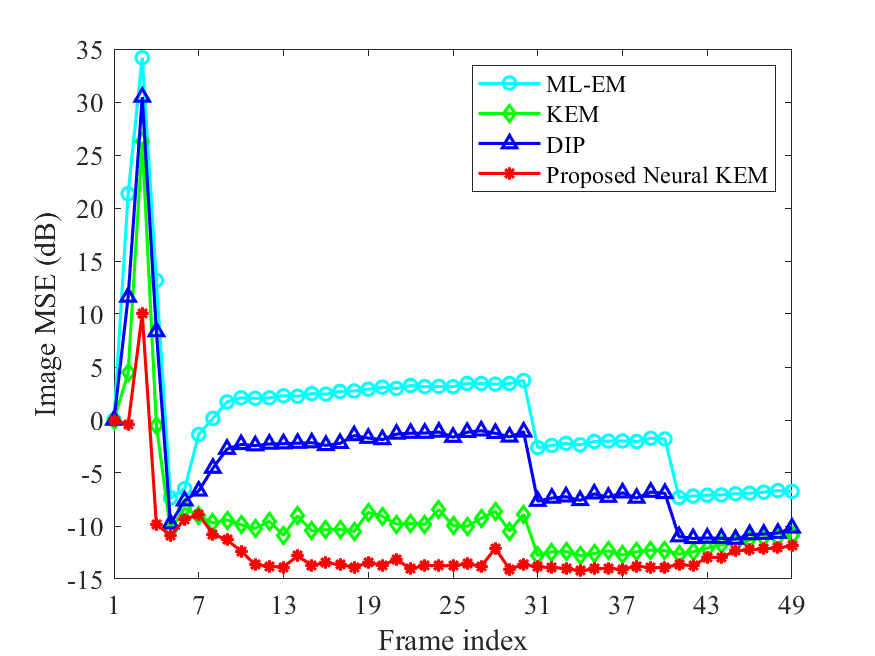}}
	\label{MSE_all}
	\subfloat[]{\includegraphics[trim=0.5cm 0cm 1cm 0cm, clip,height=1.7in]{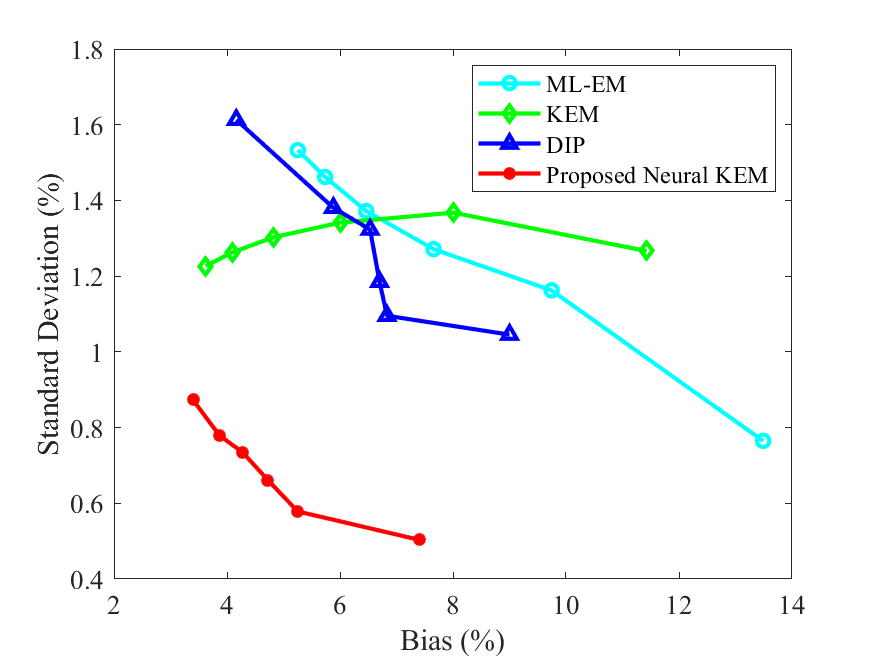}}
	\subfloat[]{\includegraphics[trim=0.5cm 0cm 1cm 0cm, 
		clip, height=1.7in]{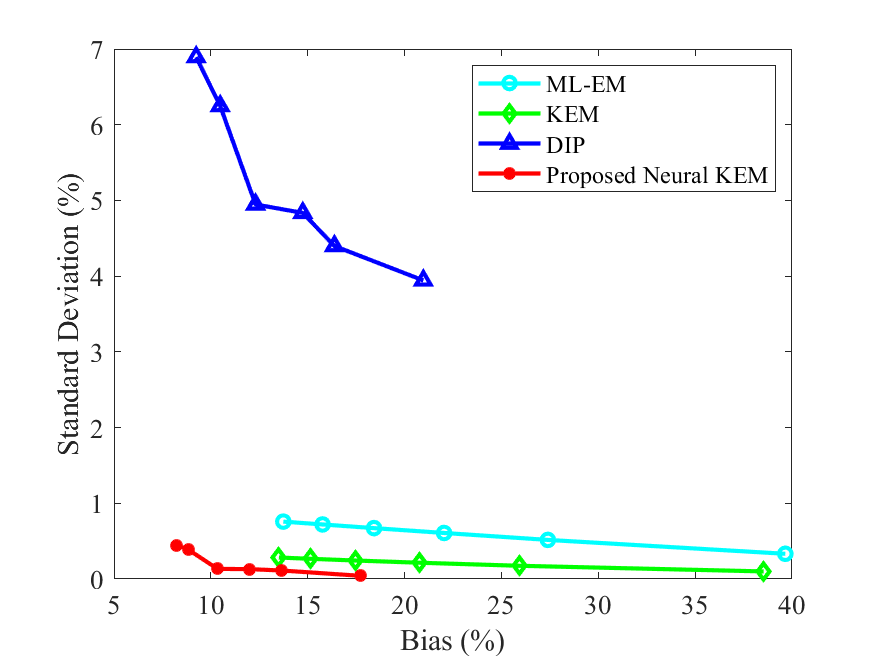}}
	\caption{\txtb{Quantitative results from the 3D simulation study. (a) Plots of image MSE across all frames; (b-c) Plots of bias-SD trade-off for ROI quantification by varying the iteration number from 10 to 60 (i.e., from rightmost to leftmost on each curve). (b) Aorta ROI in a low-count case (frame 6), (c) myocardium ROI in a high-count case (frame 49).}}
	\label{Res_3D}
\end{figure*}

\begin{figure*}[htp]
	\vspace{-0pt}
	\centering
	\subfloat[Late frame at $t = 3300-3600s$]{\includegraphics[trim=0cm 0cm 0cm 0cm, 
		clip, width=5.1in]{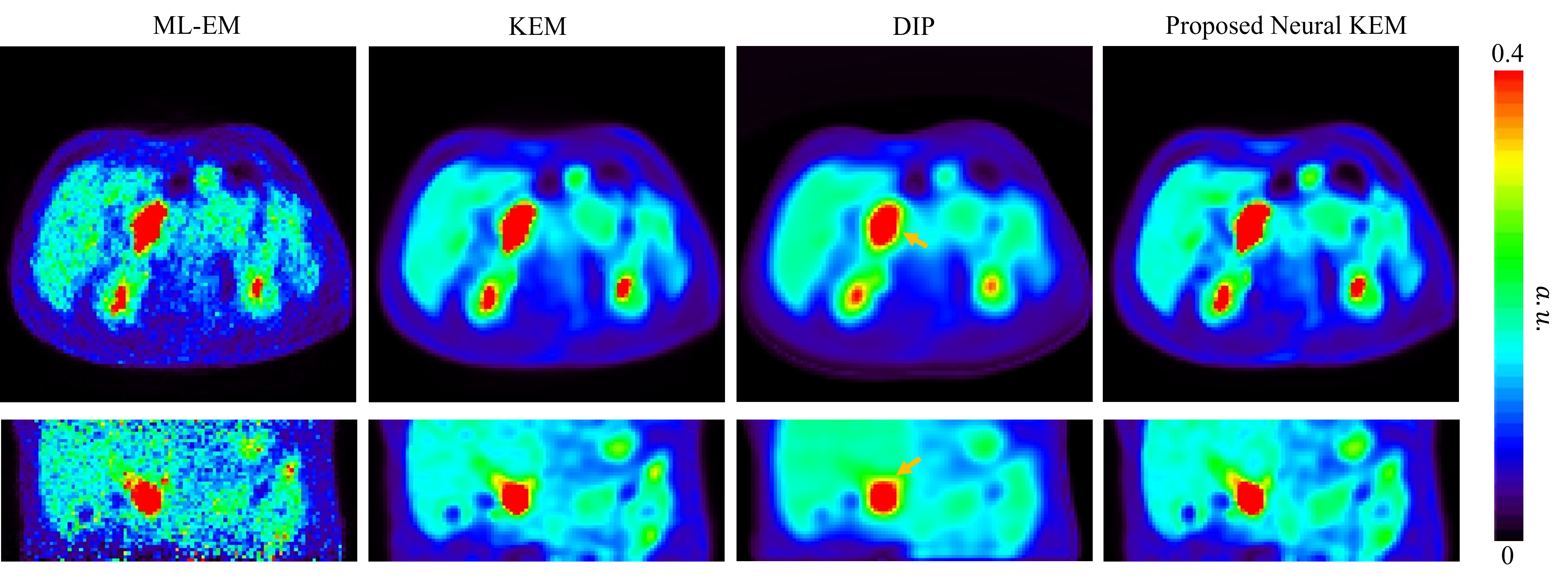}
		\label{fig_2_case}}\\
	\subfloat[Early frame at $t = 18-20s$]{\includegraphics[trim=0cm 0cm 0cm 0cm, clip,width=5.1in]{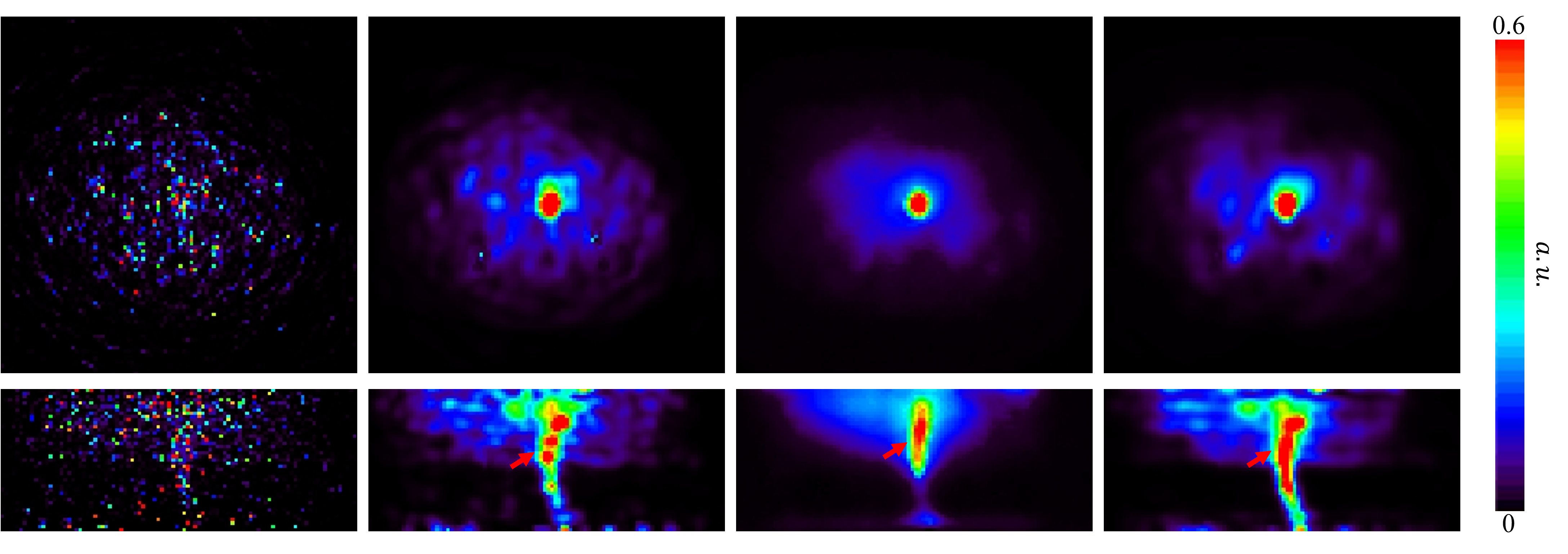}
		\label{fig_1_case}}\\
	\subfloat[Early frame at $t = 36-38s$]{\includegraphics[trim=0cm 0cm 0cm 0cm, 
		clip, width=5.1in]{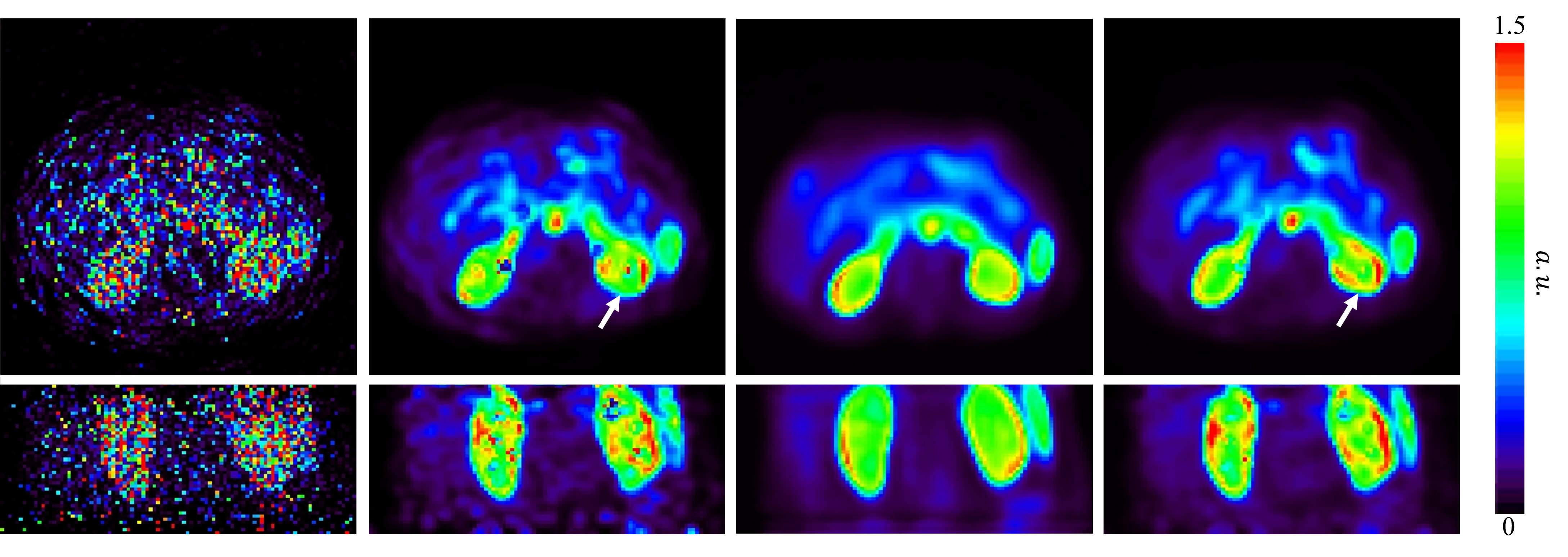}
		\label{fig_2_case}}
	\caption{3D image reconstructions of (a) a late frame at $t=3300-3600s$ and two early HTR frames at (b) $t=18-20s$ and (c) $t=36-38s$ by different methods. \txtb{Here different frames are displayed for visualizing different targets of interest, i.e.. tumor in (a), aorta in (b), and kidney in (c).} Each reconstruction is shown in the transverse and coronal views.}
	\label{real}
\end{figure*}

\section{\txtb{Validation using 3D Computer Simulation}}
\subsection{Simulation Setup}
\txtb{We also performed a fully-3D computer simulation study using an XCAT heart phantom for a GE Discovery 690 PET scanner. This scanner has 13,824 LYSO crystals, arranged in 24 ring detectors. The detection unit is composed of blocks consisting of 9$\times$6 crystals, each containing a total of 64 blocks per ring. The sinogram size is $381\times553\times288$ and the image size is $137\times137\times47$. The voxel size was $4.0\times4.0\times3.3$ mm$^3$. A one-hour dynamic $^{18}$F-FDG scan was simulated using 49 time frames: 30$\times$10s, 10$\times$60s, and 9$\times$300s. Here the framing scheme is adapted to capture the fast kinetics in the heart. The TACs of different regions were extracted from a real patient FDG PET scan to generate noise-free dynamic activity images.  The images were then forward projected to generate noise-free sinograms. No time-of-flight information was simulated. To reduce time, SimSET-based simulation was not used for scatter simulation here. Instead, scattered and random events were simulated using a 40\% uniform sinogram. Poisson noise was then generated with 1.25 billion expected events over 1 hour. Ten noisy realizations were simulated.}

\subsection{Reconstruction Methods}

\txtb{For the kernel methods, the kernel matrix $\K$ was built using four composite frames (one 5-min frame, one 15-min frame and two 20-min frames). Here compared to the 3 composite frames used for the 2D simulation study, the first 20-min was divided into two shorter composite frames to better capture the early dynamic information for reconstructing the data of higher temporal resolution. The kNN search was performed in a $9\times9\times9$ local region with $k=50$ nearest neighbors to reduce the computation time. The 3D version of U-net was used in the neural network-based methods with the 4 composite images as the network input. Similar to the 2D simulation study, 150 iterations were used for the neural-network learning step within each outer iteration. The proposed neural KEM was compared with ML-EM, regular KEM and DIP by OT using the image MSE and ROI bias-SD metrics. All the methods were run for 60 iterations starting from a uniform initial image.}

\subsection{Evaluation Results}

\txtb{Fig. \ref{3D simulation} shows the true 3D activity images and reconstructed images at iteration 60 by different reconstruction methods for frame 6 (early 10-s frame, low-count level) and frame 49 (late 5-min frame, high-count level). The proposed neural KEM achieved the lowest image MSE for both frames.}

\txtb{Fig. \ref{Res_3D}(a) shows the plots of image MSE for all frames reconstructed by different methods with 60 iterations. The two kernel-based methods (regular KEM and neural KEM) demonstrated a substantial improvement as compared to the ML-EM and DIP methods. The neural KEM was further better than the regular KEM particularly for those low-count frames.}

\txtb{Fig. \ref{Res_3D}(b) and fig. \ref{Res_3D}(c) show the trade-off between the bias and SD of different methods for ROI quantification in an aorta ROI of 3077 voxels and a myocardium ROI of  3744 voxels, respectively. The result of this 3D simulation study here is consistent with that of the 2D simulation study. The DIP method demonstrated a poor ROI bias-SD performance for the myocardial ROI quantification even though the associated image MSE was better than the MLEM. In comparison, the proposed neural KEM achieved the best trade-off  among the different reconstruction methods.}

\section{Application to Real Patient Data}

\subsection{Patient Data Acquisition}
We have further applied the neural KEM to dynamic PET imaging for a real patient dataset. A cancer patient scan was performed on the GE Discovery 690 PET/CT scanner at the UC Davis Medical Center. The PET scan started right at the injection of 10 mCi $^{18}$F-FDG and lasted 60 minutes. As our simulation studies have indicated that the neural KEM mainly benefits low-count frames, here we focus on high-temporal resolution dynamic imaging.  The one-hour data are divided into 97 time frames following the schedule $60\times2$s, $18\times10$s, $10\times60$s, and $9\times300$s. A CT scan was acquired for attenuation correction. The projection data size was $381\times553\times288$ and the image size was $192\times192\times47$. All data corrections, including normalization, attenuation correction, scatter correction and random correction, were extracted using the vendor software and included in the reconstruction process. \txtb{Similar to the 3D simulation study, four composite frames (one 5-min frame, one 15-min frame and two 20-min frames)) were used to build the kernel matrix for this high-temporal resolution reconstruction. Other algorithm settings were also the same as used for the 3D simulation study described in Section V.A.}

\txtb{For ROI analyses, three ROIs were manually drawn on the corresponding CT images in the descending aorta, the tumor, and the kidney cortex regions. The volume of the blood ROI, tumor ROI and kidney ROI is 9 cm$^3$, 12 cm$^3$, and 15 cm$^3$, respectively.}

\begin{figure*}[t]
	\vspace{-0pt}
	\centering
	\subfloat[]{\includegraphics[trim=0.0cm 0cm 1cm 0cm, clip,width=2.1in]{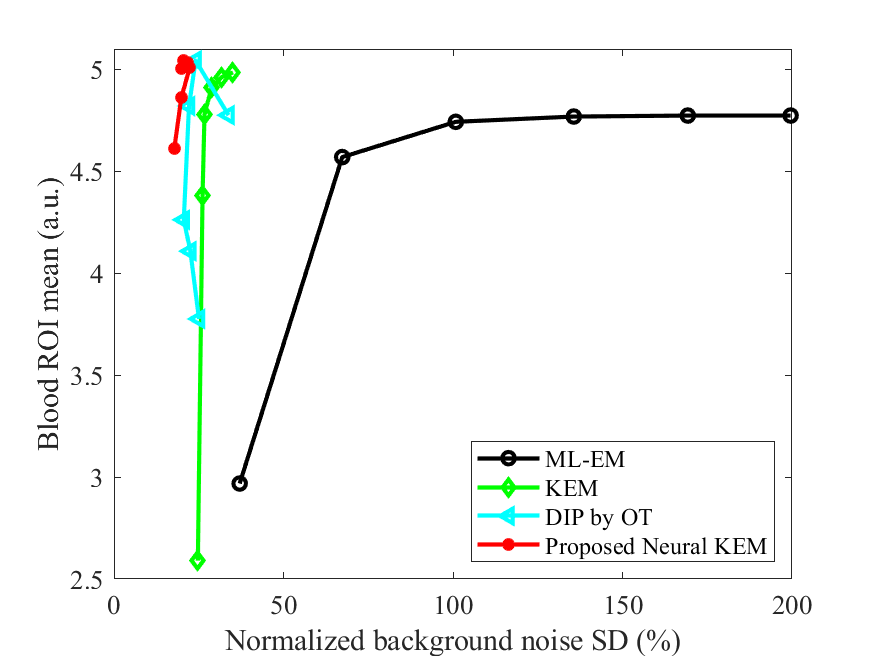}
		\label{fig_3_case}}
	\hfil
	\subfloat[]{\includegraphics[trim=0.0cm 0cm 1cm 0cm, 
		clip, width=2.1in]{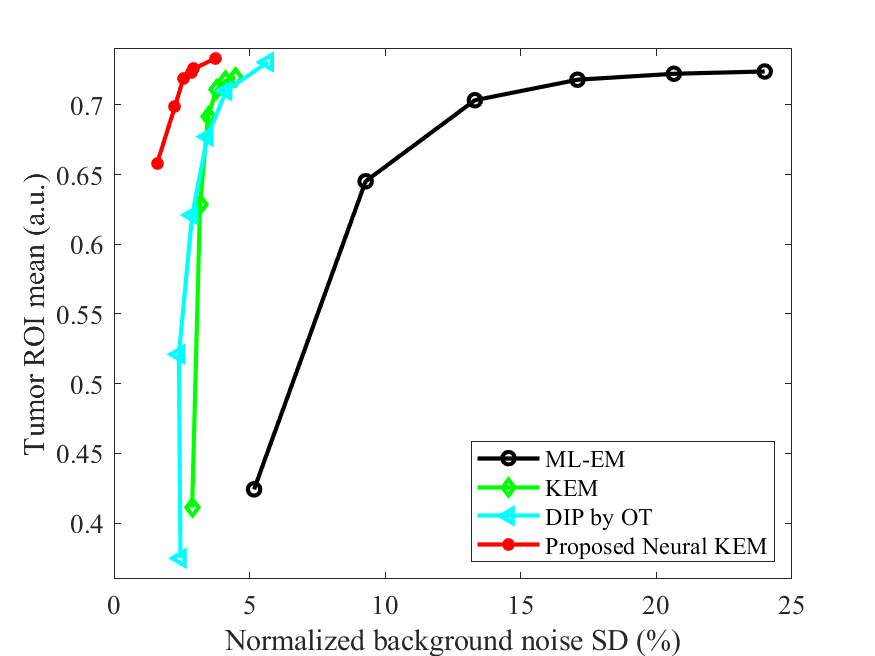}
		\label{fig_4_case}}
	\hfil
	\subfloat[]{\includegraphics[trim=0.0cm 0cm 1cm 0cm, 
		clip, width=2.1in]{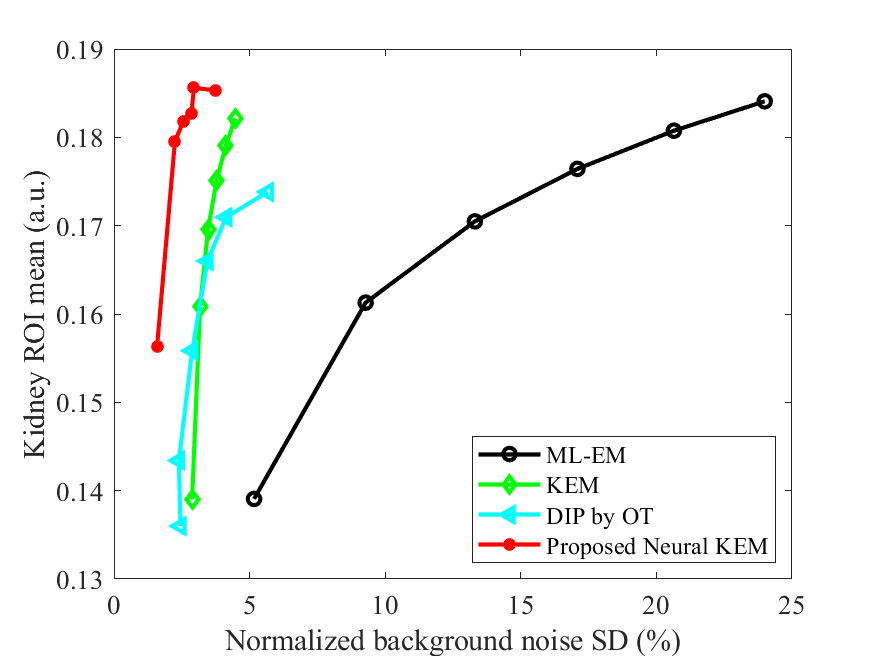}
		\label{fig_4_case}}
	\caption{Plots of ROI mean of (a) the blood region, (b) tumor, and \txtb{(c) kidney cortex} versus liver background noise by varying iteration number from 10 to 60.}
	\label{real_mean}
\end{figure*} 

\subsection{Results of Reconstructed PET Images}

Fig. \ref{real} shows the comparison of different reconstructions for a late 5-minute frame ($t=3300-3600s$) and two early 2-s frames ($t=18-20s$, $t=36-38s$). Each reconstruction is shown in the transverse and coronal views. 

For the 5-minute frame which has a relatively high count level, the ML-EM reconstruction had good contrast in the tumor region though still contained high noise in the normal liver parenchyma. The DIP by OT caused over-smoothness and demonstrated a distortion in the tumor as pointed by the arrow in Fig. \ref{real}(a). With a preserved tumor shape similar to that of the ML-EM, both the regular KEM and neural KEM suppressed noise well and  provided similar results in this high-count reconstruction, \txtb{though the latter may have a slightly higher risk of oversmoothing small targets or sharp edges for higher count data due to the additional regularization from the use of deep coefficient prior. }

\begin{figure*}[htp]
	\vspace{-0pt}
	\centering
	\subfloat[$v_b$]{\includegraphics[trim=0cm 0cm 0cm 0cm, clip,width=6.7in]{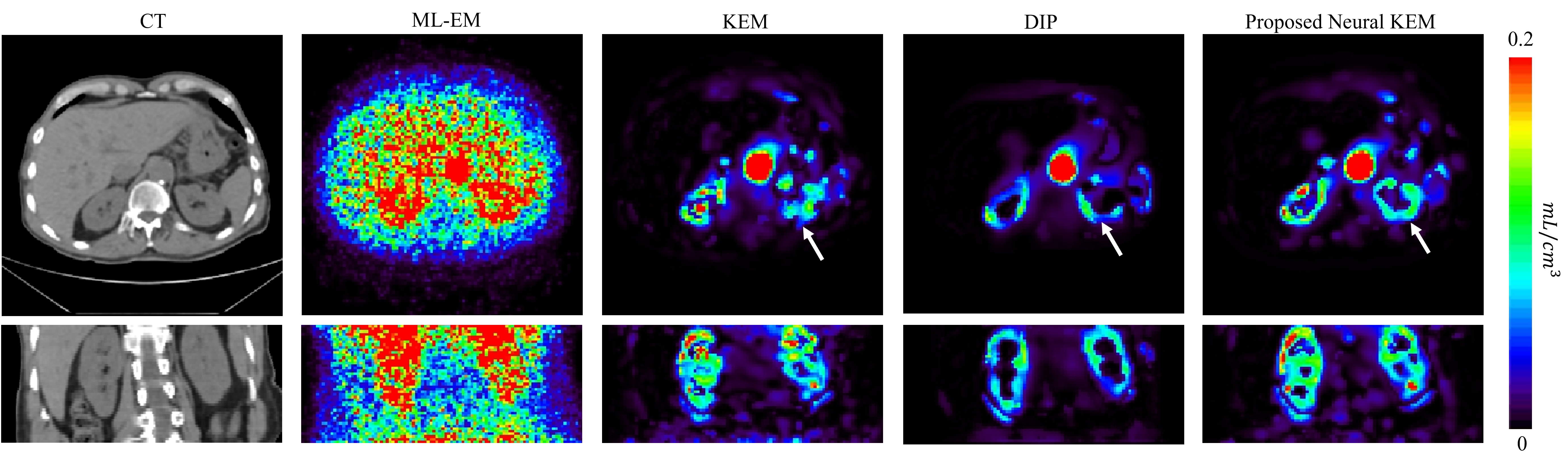}
		\label{fig_1_case}}\\
	\subfloat[$K_1$]{\includegraphics[trim=0cm 0cm 0cm 0cm, 
		clip, width=6.7in]{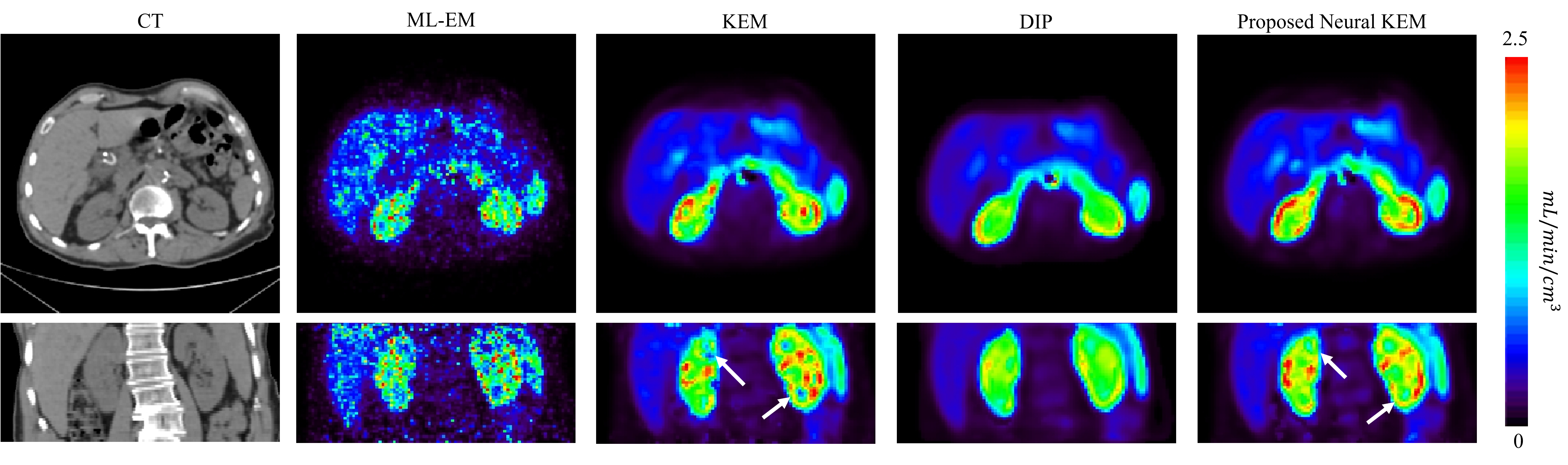}
		\label{fig_2_case}}
	\caption{Parametric images of (a) $v_b$ and (b) $K_1$ generated from the early-dynamic images reconstructed using ML-EM, KEM, DIP and the proposed neural KEM. Each image is shown in transverse and coronal views.}
	\label{Parametric image}
	\vspace{-0pt}
\end{figure*} 

For the 2-s frames, the ML-EM reconstructions were extremely noisy. The DIP by OT significantly reduced noise but tended to over-smooth the images. \txtb{In Fig. \ref{real}(b), KEM resulted in a discontinuous aorta while the proposed neural KEM showed a more natural shape. In Fig. \ref{real}(c), the neural KEM also showed a more continuous renal cortex, which may benefit parametric imaging as will be presented in the next subsection.} To sum up, the proposed neural KEM not only suppressed the noise in the background regions but also preserved structural contrast and details, \txtb{though it cannot exclude a potential risk of over-regularization similar to any other methods that includes a regularization}. Here the neural KEM and DIP by OT showed different anatomical structures in these two low-count frames. While there is no ground truth and the ML-EM was too noisy to provide a reference, the result from the high-count frame shown in Fig. \ref{real}(a) may imply the result by the neural KEM is more likely to be close to the truth of the low-count frames. 

Fig. \ref{real_mean} further shows a quantitative comparison of different methods for ROI quantification in the blood ROI in frame 15 (where the uptake in the blood reaches its maximum), and in the tumor ROI and \txtb{kidney ROI} in the last frame (55-60 min). Here the ROI mean is plotted versus normalized background noise SD by varying the iteration number from 10 to 60. The proposed neural KEM achieved the best trade-off among all the other methods.

\subsection{Demonstration for Parametric Imaging}

Parametric imaging was also performed for the dynamic images of the same subject. Because different reconstruction methods mainly make a difference for early-time frames which have a low count level (Fig. \ref{real}), here we focused on parametric imaging of early-dynamic data. A two-tissue compartment model with voxel-wise time delay estimation \cite{Wang2021} was used to generate parametric images from the early-dynamic data. \txtb{For each reconstruction method, the blood input function was derived from the descending aorta ROI.}

Fig. \ref{Parametric image} shows the parametric images of fractional blood volume $v_b$ and FDG delivery rate $K_1$ generated from the early 120s data. The CT images are also shown for reference of anatomy. The ML-EM result suffered from heavy noise. The KEM result demonstrated a significant improvement but still suffered from noise-induced artifacts. The proposed neural KEM showed more complete and regular kidney cortex structures that seem consistent with the kidney anatomy and function \cite{Ruiz2020}. \txtb{The $K_1$ by DIP was much lower in the kidney cortex region, which can be explained by the underestimation of the renal uptake in the DIP-reconstructed activity image as shown in Fig. \ref{real}(c) and Fig. \ref{real_mean}(c).}

\section{Discussion}

This work proposed an implicit regularization for improving the kernel method using deep coefficient prior and developed a neural KEM algorithm for neural-network based tomographic reconstruction. Because the loss function (\ref{surrogate}) used for the CNN learning is derived from the optimization transfer theory, the proposed neural KEM is thus guaranteed to monotonically increase the data likelihood. Compared to the ADMM used in most DIP reconstructions \cite{Gong2019, Gong2019a, Xie2020, Xie2021}, the optimization transfer algorithm does not introduce an additional hyper-parameter. The results shown in Fig. \ref{fig:5} and Fig. \ref{fig:7} indicate a more stable performance of the optimization transfer algorithm than the ADMM.

\txtb{Our studies showed mixed results for comparing DIP with the standard KEM, while DIP was reported superior over KEM in \cite{Gong2019} for MRI-guided PET image reconstruction. This is likely due to the use of different sources of image prior (MRI vs composite images of dynamic PET) and the fact that dynamic PET consists of frames of a range of count levels. The KEM or DIP alone demonstrated an instability (too noisy or too smooth) for dynamic PET image reconstruction. By combining them together, the proposed neural KEM achieved a much better performance than each individual method.}

The neural KEM in this work focused on frame-by-frame image reconstruction in the spatial domain but can be potentially extended to more general cases.  For example, a spatiotemporal kernel method \cite{Wang2019} allows both spatial and temporal correlations to be encoded in the kernel matrix. The proposed neural KEM algorithm may be combined with the spatiotemporal kernel method to further improve the dynamic image reconstruction of high-temporal resolution data. In addition, the construction of a spatial kernel itself can also be modified by using \txtb{a different kernel function, e.g., using a pre-defined wavelet representation \cite{Ashouri2021} or a neural-network representation \cite{Li2021}. The kernel construction can be further trained using deep learning as demonstrated in our recent work \cite{Li2022}. It is worth noting that all these methods are aimed at improving $\K$ and are therefore complementary to the proposed neural KEM which improves $\alp$ in (\ref{kernel model}). Our preliminary results from computer simulation (not shown) have suggested that the observed benefit of using the deep coefficient prior on $\alp$ in this paper can be also transferred into those methods that use a modified or trained kernel. A detailed study will be reported in our future work.}

Compared to the standard KEM,  the neural KEM introduces a nonlinear step (i.e., the neural network learning step), which brings the image quality improvement but adds extra computational cost. \txtb{For example, the standard KEM took 30 minutes while the neural KEM took 70 minutes for reconstructing one frame in the real data study.} A potential way to reduce the computational burden is to accelerate the speed of the neural network learning step either by improving the learning algorithm or by using pretraining for a better initial.



In this study, we demonstrated the performance of the proposed algorithm on conventional PET scanners. The recent advent of total-body PET scanners (e.g., \cite{Cherry2017, Badawi2019, Spencer2021, Karp2018, Pantel2020}) has made it even more feasible to pursue  low-dose dynamic imaging and high-temporal resolution dynamic imaging, especially for the entire body simultaneously. Our future work will also implement and evaluate the proposed neural KEM on total-body PET for parametric imaging.

\section{Conclusion}

In this paper, we have developed a neural KEM algorithm that combines the kernel method with deep coefficient prior. The algorithm is enabled by optimization transfer, leading to an easy-to-implement modularized implementation. Computer simulations and real patient data have demonstrated the improvement of the neural KEM over conventional KEM and DIP methods for dynamic PET imaging. 

\section*{Acknowledgment}

We thank Dr. Benjamin Spencer for assistance in patient data acquisition and Yiran Wang for assistance in generating the parametric images.

\end{document}